\documentclass[prb,twocolumn,showpacs,preprintnumbers,amsmath,amssymb,superscriptaddress]{revtex4}
\usepackage{graphicx}
\usepackage{dcolumn}% Align table columns on decimal point
\usepackage{bm}% bold math

\usepackage{epstopdf}

\usepackage{color}
\newcommand{\blue}{\color{black}}
\newcommand{\red}{\color{black}}
\newcommand{\green}{\color{black}}
\newcommand{\cyan}{\color{black}}

\begin{document}

\title{Crystal dynamics and thermal properties of neptunium dioxide}

\author{P. Maldonado}\affiliation{Department of Physics and Astronomy, Uppsala University, Box 516, S-75120 Uppsala, Sweden}
\author{L. Paolasini}\affiliation{European Synchrotron Radiation Facility (ESRF), B.P.\,220, F-38043 Grenoble, France}
\author{P. M. Oppeneer}\affiliation{Department of Physics and Astronomy, Uppsala University, Box 516, S-75120 Uppsala, Sweden}
\author{T. R. Forrest}\affiliation{European Synchrotron Radiation Facility (ESRF), B.P.\,220, F-38043 Grenoble, France}
\author{A. Prodi}\affiliation{Consiglio Nazionale delle Ricerche, Istituto di Struttura della Materia, Area della Ricerca Roma 1, via Salaria km 29.300, Montelibretti, Italy}
\author{N. Magnani}\affiliation{European Commission, Joint Research Centre (JRC), Institute for Transuranium Elements (ITU), Postfach 2340, D-76125 Karlsruhe, Germany}
\author{A. Bosak}\affiliation{European Synchrotron Radiation Facility (ESRF), B.P.\,220, F-38043 Grenoble, France}
\author{G.~H. Lander}\affiliation{European Commission, Joint Research Centre (JRC), Institute for Transuranium Elements (ITU), Postfach 2340, D-76125 Karlsruhe, Germany}
\author{R. Caciuffo}\affiliation{European Commission, Joint Research Centre (JRC), Institute for Transuranium Elements (ITU), Postfach 2340, D-76125 Karlsruhe, Germany}

\date{\today}

\begin{abstract}
We report an experimental and theoretical investigation of the lattice dynamics and thermal properties of the actinide dioxide
NpO$_2$. The energy-wavevector dispersion relation for normal modes of vibration
propagating along the $[001]$, $[110]$, and $[111]$ high-symmetry lines in  NpO$_2$  at
room temperature has been determined by measuring the coherent one-phonon scattering
of X-rays from a $\sim$1.2 mg single-crystal specimen, the largest available single crystal for this compound. The results are compared against \textit{ab initio} phonon dispersion{\red s}
%simulations
computed within the first-principle{\red s} density functional theory in the generalized gradient approximation plus Hubbard $U$ correlation (GGA+$U$) approach, taking into account third-order anharmonicity effects in the quasiharmonic approximation. Good agreement with the experiment is obtained for
calculations with an on-site Coulomb parameter $U = 4$ eV and Hund's exchange $J= 0.6$ eV in line with previous electronic structure calculations. We further compute the thermal expansion, heat capacity, thermal conductivity, phonon linewidth, and thermal phonon softening, and compare with available experiments. The theoretical and measured heat capacities are in close agreement with another. About {\red 27}\% of the calculated thermal conductivity is due to phonons with energy higher than 25 meV ($\sim$ 6 THz ), suggesting an important role of high-energy optical phonons
in the heat transport. The simulated thermal expansion reproduces well the experimental data
%only
up to about 1000 K, indicating a failure of the quasiharmonic approximation above this limit.
\end{abstract}


\maketitle

\section{Introduction}
Thermal neutrons with energy matching that of atomic dynamics in crystalline materials generally have a momentum comparable in size to the first Brillouin zone of the reciprocal space lattice. This makes energy- and momentum-resolved inelastic neutron scattering (INS) a powerful tool for measuring phonon dispersion in crystals, as first demonstrated by Brockhouse {\red and collaborators} \cite{brockhouse58}. However, neutron techniques are intensity-limited and become unpractical when crystal samples of sufficient size are not available. This is the case of NpO$_{2}$, for which only single crystals smaller than $\sim$10$^{-4}$ cm$^{3}$ have been obtained. NpO$_{2}$ has been the object of considerable attention because of its peculiar physical properties, but the lack of large single crystals has prevented any experimental determination of its lattice dynamics. On the contrary, the phonon dispersions for the isomorphous uranium dioxide have been extensively investigated by INS from the mid 1960s \cite{dolling65} to the present day \cite{pang13,pang14}.

Since phonons largely determine the thermal conductivity of actinide dioxides, the backbone of nuclear reactor fuels, measuring their dispersion yields important information of technological interest \cite{pang13}.
%On the other hand,
{\red In addition,} phonons are supposed to be an important ingredient of the low-temperature physics of these compounds. This is the case of UO$_{2}$, whose ground state and low-energy collective excitations are governed by the interplay between crystal field (CF) interactions \cite{magnani05}, two-ion multipolar interactions of purely electronic origin, and magnetoelastic coupling between the uranium magnetic moment and the oxygen displacements \cite{caciuffo99,carretta10,caciuffo11}. Electron-phonon interactions are also supposed to generate a crystal field-phonon bound state in NpO$_{2}$ \cite{amoretti92} and to play an important role in the stabilization of the multipolar order. The low-temperature phase of NpO$_{2}$ has indeed  provided the first example of hidden order in actinide systems \cite{caciuffo87,santini09}. The large anomalies observed\cite{westrum53,osborne53} at $T_{\rm o} = 25$ K in specific-heat and magnetic-susceptibility measurements  have eventually been interpreted as signature of a phase transition where the magnetic triakontadipole of $\Gamma_{5}$ symmetry plays the role of primary order parameter (OP) \cite{santini06,magnani08}, whilst electric quadrupoles are induced as secondary OP and dipolar magnetic moments are quenched \cite{caciuffo03,suzuki10}. As the order is longitudinal 3-k, the overall cubic symmetry of the lattice is preserved  \cite{paixao02}.

Actinide dioxides are also good benchmark compounds to test and refine existing theoretical and computational methods. Very few attempts to calculate \textit{ab initio} the physical properties of NpO$_{2}$ have been published \cite{maehira07,prodan07,wang10,suzuki13}, in part because very scarce single-crystal data are available.
To fill this information gap, we used non-resonant, meV-resolved inelastic X-ray scattering (IXS) to measure the phonon energy spectrum $\hbar \omega_{j}(\bm{q})$ (where $j$ is the branch polarization index and $\bm{q}$ the wavevector) of NpO$_{2}$ along the high-symmetry directions [$\Gamma-X$], [$X-K-\Gamma$], and [$\Gamma-L$] of the \textit{fcc} Brillouin zone. Although not as widespread as INS, IXS is a mature technique available at several third-generation synchrotron radiation sources, which allows one to study collective vibrational dynamics using micro-gram scale crystals \cite{krisch07}.
{\red Recently, IXS was used to measure\cite{manley12} the phonon density of states of Ga-doped PuO$_2$.}
Compared with neutrons, IXS offers advantages linked to a very low intrinsic background, an energy resolution decoupled from energy transfer, and an energy-independent momentum transfer. However, as the scattering cross-section is proportional to the square of the atomic number, observing the contributions of light atoms to the vibrational spectra by IXS is challenging.
The {\red here-reported} experiment was performed with the ID28 spectrometer of the European Synchrotron Radiation Facility (ESRF) in Grenoble, France. The measured dispersions are in excellent agreement with the phonon spectrum calculated by means of the
{\red GGA+$U$ electronic structure approach in combination with the quasiharmonic approximation.}
%quasiharmonic local density approximation (GGA)+$U$ approach.

The simulated phonon dispersions have been used to calculate the vibrational contribution to several thermodynamical quantities, namely, thermal expansion, heat capacity, thermal conductivity, phonon linewidth, and thermal phonon softening. The results are in very good agreement with available experimental data and suggest an important contribution of high-energy optical phonons
in the heat transport. A departure between experimental and simulated thermal expansion curves above $\sim$ 1000 K shows that the use of the quasiharmonic approximation at higher temperatures may result in wrong estimates of some thermodynamical observables.

\section{Experimental methodology}
Due to the contamination risk generated by the radiotoxicity of the neptunium element, all operations of preparation and encapsulation have been carried out in shielded gloveboxes under inert nitrogen atmosphere following well-established safety procedures. The experiment was carried out at room temperature on a {\green high quality} single crystal of NpO$_{2}$ using the ID28 beamline with an incident energy E$_{i}$ = 17.794 keV, afforded by a flat Si (999) perfect crystal back-scattering monochromator. The monochromator was temperature controlled in the mK region by a high-precision platinum thermometer bridge in closed-loop operation with a controlled heater unit.
The used analyser was formed by 12000 Si crystals of 0.6$\times$0.6$\times$3 mm$^{3}$ size glued onto a spherical silicon substrate and thermally stabilized to 6$\times$10$^{-4}$ K. {\green This configuration gave a constant energy resolution of 3 meV.} The chosen incident energy is just above the Np $L_{3}$ absorption edge energy ($E_{L_{3}}$ = 17.610 keV), but below both the $L_{1}$ and $L_{2}$ edges ($E_{L_{1}}$ = 22.427 keV, $E_{L_{2}}$ = 21.601 keV). As a result, a severe sample photo-absorption is present, and the use of transmission scattering geometry must be avoided. This energy represents also a good compromise between the energy resolution and the optimization of the photon flux. {\green In order to reduce the beam path on the sample surface, an Ru/B$_{4}$ mirror multilayer focusing configuration was used to produce a beam
spot size of 30$\times$80 $\mu$m$^{2}$ on the sample surface.}
The crystal of dimension of 0.78$\times$0.56$\times$0.25 mm$^{3}$ ($\sim$1.2 mg) was oriented with the specular direction along the $\langle001\rangle$ axis, with the $\langle110\rangle$ axis in the scattering plane. The sample  was encapsulated between two diamond slabs of 0.5$\times$5$\times$5 mm$^{3}$ at the Institute for Transuranium Elements in Karlsruhe. The diamond slabs where oriented with their $\langle110\rangle$ axis closely parallel to the Np $\langle001\rangle$ direction. Diamond phonon groups (mainly acoustic) were detected around the $(400)$ and $(300)$ NpO$_{2}$ Brillouin zones (BZs) centers and were distinguishable from the NpO$_{2}$ phonons because of their very steep dispersion curves. However, due to the weakness of the optic phonons of NpO$_{2}$, these contaminated BZs were avoided.

\section{Theoretical methodology}

\subsection{Harmonic and quasiharmonic lattice dynamics and lattice thermal conductivity}\label{sec-3A}

In the harmonic phonon approximation thermodynamic properties are described through the Helmholtz free energy \cite{togo10}
\begin{equation}
F(T,V) =
{\frac{1}{2}\sum_{\bm{q},\nu}\hbar\omega (\bm{q},\nu)+k_{B}T\sum_{\bm{q},\nu}\ln \left[1-\textrm{e}^{\frac{-\hbar\omega (\bm{q},\nu)}{k_{{\scriptscriptstyle B}} T}}\right]},
\label{eq1}
\end{equation}
where $ \hbar \omega (\bm{q},\nu)$ is the energy of the phonon mode $\nu$ at the wavevector $\bm{q}$ in the reciprocal space, $V$ and $T$ are the volume and temperature of the system, respectively, and $k_{B}$ is the Boltzmann constant.
The first term on the right hand side accounts for the phonon mode zero-point vibrational energy and the second term is the contribution of each mode to the free energy due to thermal occupation of the phonon
energy levels. It is convenient to rewrite Eq.\ (\ref{eq1}) in terms of the phonon density of states $g(\omega ,V)$ as
\begin{equation}
F(T,V)= \displaystyle{\int_{0}^{\infty} \!\! d\omega \, g(\omega,V )\left(\frac{\hbar\omega}{2}+k_{B}T \ln \left[1-\textrm{e}^{\frac{-\hbar\omega}{k_{\tiny B}T}}\right]\right)},
\label{eq2}
\end{equation}
where now the integral is over the phonon frequencies.

The Gibbs free energy $G$, from which thermodynamic properties at a temperature $T$ can be computed, is obtained by minimizing the function $F(V,T)$ for a given (constant) pressure $p$
\begin{equation}
G(T,p)=\min_{V}\left[ U(V)+F(T,V)+pV \right]
\label{eq3}
\end{equation}
with $U (V)$ being the volume-dependent electronic total energy.

It is important to stress that within a harmonic potential the normal phonon mode's frequencies of a crystal are unaffected by a change in volume, and as a consequence the system does not render any thermal expansion. This is due to the fact that in a harmonic model the only temperature dependence is in the phonon occupation numbers and therefore the equilibrium volume cannot vary with temperature at fixed pressure. However, in real crystals the harmonic approximation is not exact and the normal phonon modes depend on the volume of the crystal due to anharmonic terms. A way to take into account these anharmonicities, and thus be able to evaluate the thermal expansion, is to use the quasi-harmonic approximation (QHA). This approximation assumes that phonon frequencies are volume dependent, but that, at a given volume, the interatomic forces are harmonic. Even though this assumption is only valid for small anharmonic perturbations, in practice it is a reasonable accurate approximation for temperatures below half the melting temperature. To conduct QHA calculations it is needed to compute the normal phonon modes for different crystal volumes around the equilibrium volume, and calculate the Helmholtz free energy for each as a function of temperature. Thus, the Gibbs free energy can be  obtained  at arbitrary temperatures by fitting the Helmholtz free energy as a function of  volume to a pertinent equation of state. Subsequently, different thermodynamic properties (such as the linear or volumetric expansion coefficients, the bulk modulus, and the constant-pressure heat capacity) can be calculated.

From classical kinetic theory the rate of change in the phonon distribution function can be determined via the Boltzmann transport equation,
from which the thermal conductivity $\kappa$ can be calculated, either by using the relaxation time approximation (RTA) or from a full solution (FS) of the Boltzmann transport equation \cite{togo15}. These thermal {\blue conductivity tensors} are given by
\begin{eqnarray}
\kappa_{_{RTA}}^{ij} &=& \sum_{\lambda} C_{V,\lambda}{v}_{\lambda}^i  {v}_{\lambda}^j \, \tau_{\lambda},  \label{eq-RTA}\\
\kappa_{_{FS}}^{ij} &=& \frac{\hbar^2}{4k_{B}T^2NV} \times \nonumber \\
 & & \sum_{\lambda\lambda^{\prime}}\frac{\omega_{\lambda}v_{\lambda}^i}{\sinh(\frac{\hbar\omega_{\lambda}}{2k_{B}T})}\frac{\omega_{\lambda^{\prime}}v_{\lambda^{\prime}}^j}{\sinh(\frac{\hbar\omega_{\lambda^{\prime}}}{2k_{B}T})}(\Omega^{\sim 1})_{\lambda\lambda^{\prime}},
\end{eqnarray}
%
%{\red check vector form}
where {\blue $i,~j$ denote the Cartesian coordinates,} $\lambda$ stands for $\bm{q},\nu$ and $\Omega^{\sim 1}$ denotes the Moore-Penrose inverse \cite{adi03} of the collision matrix $\Omega$ that is given by \cite{chaput13}
\begin{eqnarray}
\Omega_{\lambda\lambda^{\prime}}&=&\delta_{\lambda\lambda^{\prime}}\frac{q}{\tau_{\lambda}}+\frac{\pi}{\hbar^2}
\sum_{\lambda^{\prime\prime}}|\Psi_{\lambda\lambda^{\prime}\lambda^{\prime\prime}}|^2\frac{1}{\sinh(\frac{\hbar\omega_{\lambda^{\prime\prime}}}{2k_{B}T})}
\times \nonumber\\
&&\left[ \delta(\omega_{\lambda}-\omega_{\lambda^{\prime}}-\omega_{\lambda^{\prime\prime}})
+\delta(\omega_{\lambda}+\omega_{\lambda^{\prime}}-\omega_{\lambda^{\prime\prime}}) \right. \nonumber \\
&& \left. +\delta(\omega_{\lambda}-\omega_{\lambda^{\prime}}+\omega_{\lambda^{\prime\prime}})\right] .
\end{eqnarray}
Here $\bm{v}_{\bm{q}\nu}$ and $C_{V,\bm{q}\nu}$ are the group velocity and constant-volume heat capacity of the phonon mode $\nu$ at the $\bm{q}$-point, respectively, $\Psi_{\lambda \lambda^{\prime} \lambda^{\prime\prime}}$ is the strength of interaction between the three phonons $\lambda$, $\lambda^{\prime}$ and $\lambda^{\prime\prime}$ involved in the scattering \cite{togo15}, and $\tau_{\bm{q}\nu}$ is the mode lifetime. The phonon {\red mode} lifetime $\tau_{\bm{q}\nu}$ is calculated from  the imaginary part of the self-energy, or the phonon linewidth, $\Gamma_{\bm{q}\nu}$, via \cite{togo15}
\begin{equation}
\tau_{\bm{q}\nu}=\frac{1}{2\Gamma_{\bm{q}\nu}}.
\end{equation}

The phonon linewidths are determined using many-body perturbation theory in a third-order anharmonic Hamiltonian which only considers {\red up to} three-phonon scattering \cite{togo15}. With these considerations, its computation reduces to the knowledge of the third-order anharmonic interatomic force constants
that can be determined from density functional theory calculations.

\subsection{First-principles modelling and lattice dynamics calculations}

The electronic structure calculations were carried out using the
Vienna \textit{Ab-initio} Simulation Package (\textsc{vasp}) \cite{kresse96}, with
the generalized gradient approximation (GGA){\red \cite{perdew96}} as the DFT exchange-correlation functional, as well as with its extension to treat strongly correlated electrons, DFT with an additional Hubbard $U$ term (DFT+$U$). Within the GGA+$U$ approach, the Hubbard and exchange parameters, $U$ and $J$, respectively, are introduced to account
for the strong on-site Coulomb correlations between the neptunium $5f$
electrons. This helps to remove the self-interaction error and
improves the description of correlation effects in the open
$5f$ shell. We have chosen a Hubbard $U$ value of 4.0 eV and an exchange parameter $J$ value of 0.6 eV, which have
been shown to provide a good description of the NpO$_2$ system \cite{suzuki10,modin15} {\red and are consistent with accepted values for UO$_2$ (see, e.g.,
Refs.\ \onlinecite{dudarev97} and \onlinecite{yun11})}. Projector augmented-wave pseudopotentials were used with an energy cutoff of 500 eV for the plane-wave basis, which was sufficient to converge the total energy for a given $k$-point sampling. The Brillouin zone integrations of the employed simulation cells {\red consisting of}
%(see below),
2$\times$2$\times$2 supercells and 4$\times$4$\times$4 supercells, were performed on a special
$k$-point mesh generated by 17$\times$17$\times$17, 7$\times$7$\times$7, and 3$\times$3$\times$3  $\Gamma$-centered  Monkhorst Pack $k$-point grids respectively. The electronic minimization algorithm
used for static total-energy calculations was a blocked Davidson algorithm.

At low temperatures below $T_{\rm o} = 25$ K, an exotic multipolar ordered phase is formed in NpO$_2$ which has been the topic of many investigations, {\red see} Ref.\ [\onlinecite{santini09}] and refs.\ therein. In the present study we focus on the lattice dynamics at elevated temperatures. Therefore we choose to describe the system as an antiferromagnetic Mott insulator. After optimization of the $fcc$ fluorite structure of NpO$_2$ we find a lattice constant of 5.497 \AA, which is somewhat larger than the experimental value of 5.42 \AA. Although the GGA+$U$ method provides a slightly overestimated volume of the system, it has been shown that, {\red for actinide dioxides,} it provides a better description of the Mott gap and magnetic moments than the local density approximation (LDA) +$U$ \cite{wang10,pang14,modin15}. We find for these quantities {\red for NpO$_2$} the value{\red s} of 2.35 eV and 3.1 $\mu_B$, respectively, in agreement with previous work on NpO$_2$.  Thus, our results provide a satisfactory qualitative description of the electronic structure, and provide a good starting point for the simulation of the vibrational spectra.

Our phonon calculations were performed for NpO$_{2}$ through the finite displacement
method using the open-source package \textsc{phonopy} \cite{togo09} with \textsc{vasp} employed as the {\red density functional method}
to obtain the pairwise and cubic interatomic force constants. To evaluate these, we employed supercells consisting of 4$\times$4$\times$4 primitive cells (192 atoms) and 2$\times$2$\times$2 supercells (24 atoms). Due to the high symmetry of the cubic system only 2 and 128  sets of frozen phonon structures were needed to calculate the dynamical matrices and the phonon linewidths. During postprocessing, the phonon frequencies and lifetimes were sampled on a 50$\times$50$\times$50, and 21$\times$21$\times$21 $q$-{\red point} mesh, respectively.

For the quasiharmonic calculations, additional finite-displacement calculations were performed on simulation cells at
approximately $\pm 10\%$ of the equilibrium volume in steps of $1\%$. Due to the fact that the largest five volumes were found to correspond to temperatures outside the validity range of the QHA, these structures were omitted in the calculation of the thermal properties. Otherwise, wrong bulk modulus and thermal expansion values would be found. In addition and to correctly include the long-range macroscopic electric field generated by collective ionic motions near the $\Gamma$ point, we have added a non-analytical term to the dynamical matrix \cite{pick70}.

{\red In contrast to the {\blue DFT-}GGA methodology, the GGA+$U$ functional can lead to convergence to local minima of the total energy and not the sought global minimum. This difficulty may  lead to incorrect phonon results---these being based on total energy calculations---such as very dispersive optical phonon branches {\blue (cf.\ Ref.\ \onlinecite{sanati11})}. To avoid this problem we have used the so-called occupation matrix control technique \cite{dorado09} in our calculations.

Lastly, the spin-orbit interaction is known to be large in actinide materials and it can influence the properties of actinide materials such as the equilibrium lattice parameters \cite{kunes01}. However, for the insulating actinide dioxides it has been found previously that the influence of the spin-orbit coupling (SOC) is  small for the quite localized $5f$ states \cite{boettger00,prodan06,petit10}. Our calculations for NpO$_2$, presented below, are consistent with this observation. For most of our phonon calculations we have therefore not included the SOC, but we have evaluated the influence of the SOC by including it in a set of calculations {\blue (shown below).}
}

\section{Results}

\subsection{Phonon dispersions}

\begin{figure}
\includegraphics[width=0.9\columnwidth]{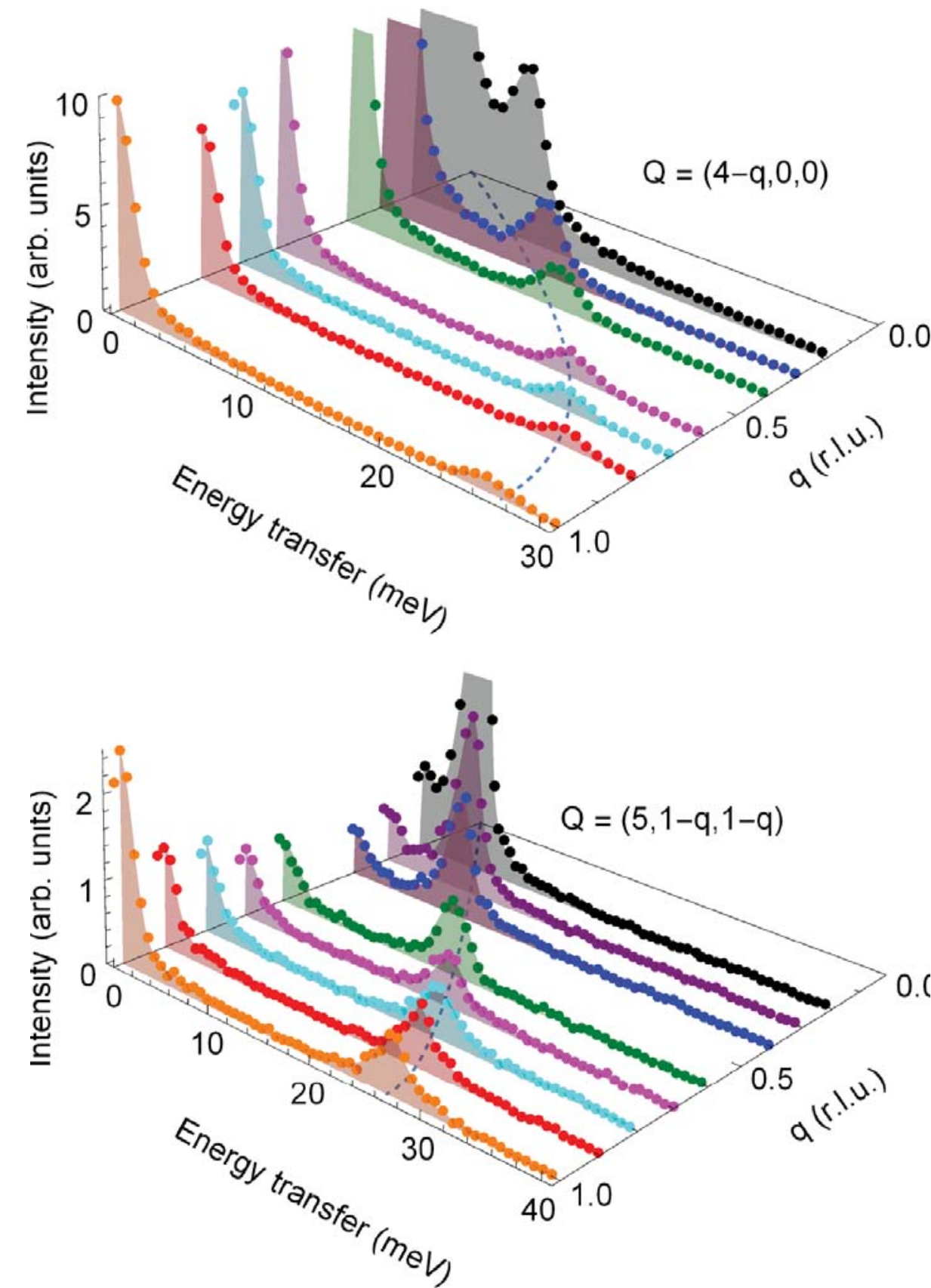}
\caption{{\red (Color online)} Dispersion of {\blue longitudinal}  acoustic phonons propagating along the {\blue $[100]$} direction ({\red top} panel) and {\blue transverse}  acoustic phonons propagating along the $[011]$ direction ({\red bottom} panel). The data have been collected at the scattering vector $\bm{Q} = \hbar (\bm{k}_{i}-\bm{k}_{f})$ specified in each panel, $\bm{k}_{i}$ and $\bm{k}_{f}$ being the wavevector of incident and scattered photons, respectively.
%{\red if we have this in one column we have no problem placing the figs.!}
\label{3D}}
\end{figure}

\begin{figure}
\includegraphics[width=0.75\columnwidth]{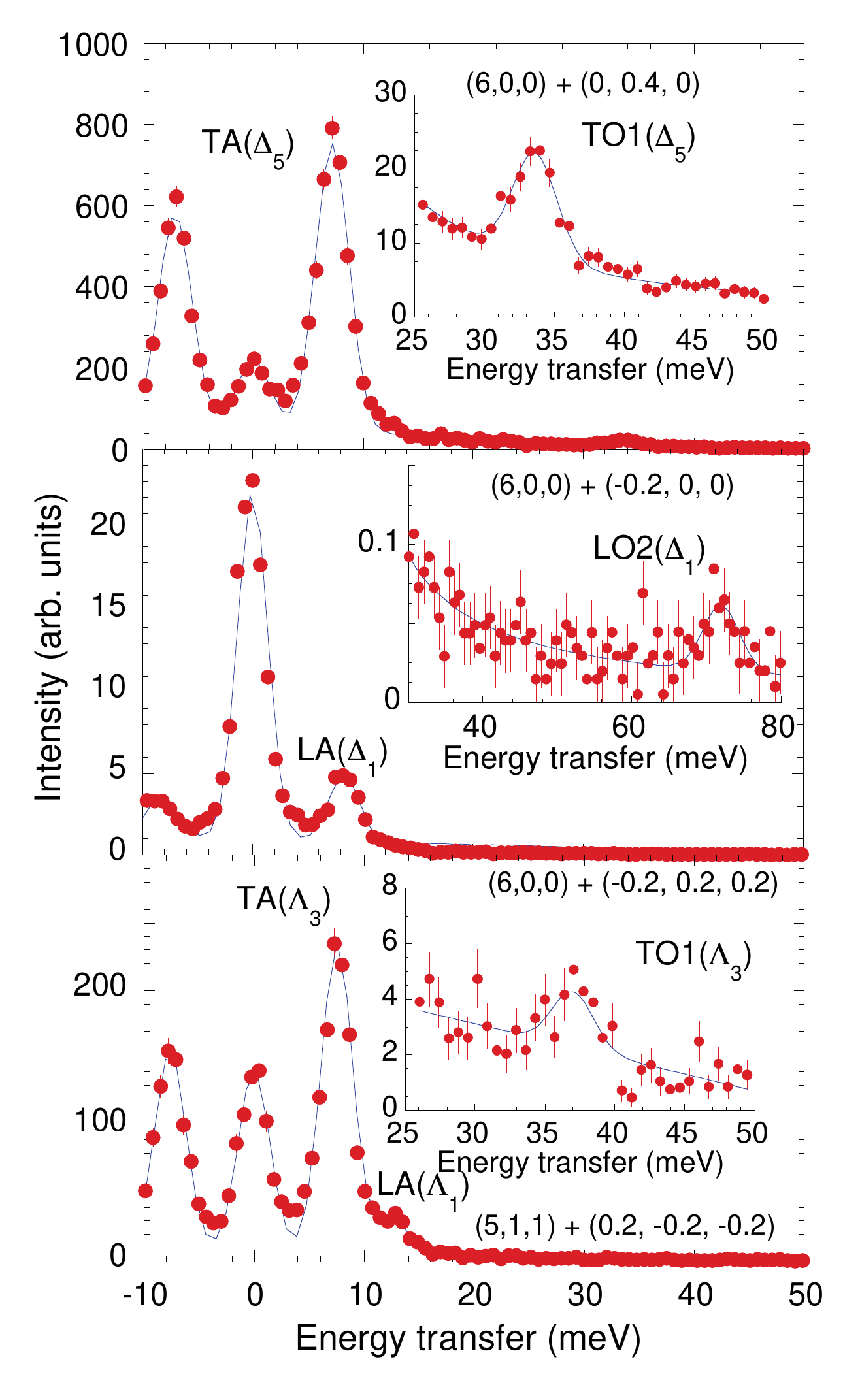}
\caption{(Color online) Experimental IXS spectra collected for representative phonon branches of NpO$_{2}$ at constant reciprocal space positions $\bm{Q}$ indicated by the labels. Optic phonon branches TO1 and LO2 have a peak intensity weaker by one or two orders of magnitude than acoustic branches. Intensity data for each panel and corresponding inset are scaled by the same factor.
\label{scans}}
\end{figure}

The acoustic phonons in NpO$_{2}$ were easily detected by IXS in all the main crystallographic directions, and a complete collection of phonon groups belonging to transverse and longitudinal acoustic branches was performed. As an example, Fig.\ \ref{3D} shows the data collected along the $[011]$ direction off the $(5,1,1)$ reciprocal lattice point (transverse acoustic branch) and along the $[100]$ direction off the $(4,0,0)$ BZ center (longitudinal acoustic branch). The variation of the inelastic peak intensity within the BZ is in agreement with the structure factors obtained from our first-principle{\red s} calculations of the vibrational spectra. Satisfactory statistics for acoustic phonon groups were obtained with counting times shorter than 30 seconds per point. On the contrary, optic phonons are much weaker (as they arise {\red mainly} from oxygen vibration modes); because of this, counting times up to three minutes per point were adopted and data were collected at a series of points $\bm{Q}$ = $\bm{q}$+$\bm{G}$ around different reciprocal lattice vectors $\bm{G}$. Inelastic structure factor simulations based on the first-principle{\red s} calculations were used as a guide in determining the most appropriate points in reciprocal space at which to conduct the measurements \cite{elliott67}. Typical phonon groups belonging to different optic and acoustic branches are shown in Fig.~\ref{scans}.
%{\red LO1 not observed, has very small metal contribution ..}

\begin{figure*}
\includegraphics[width=16.0cm]{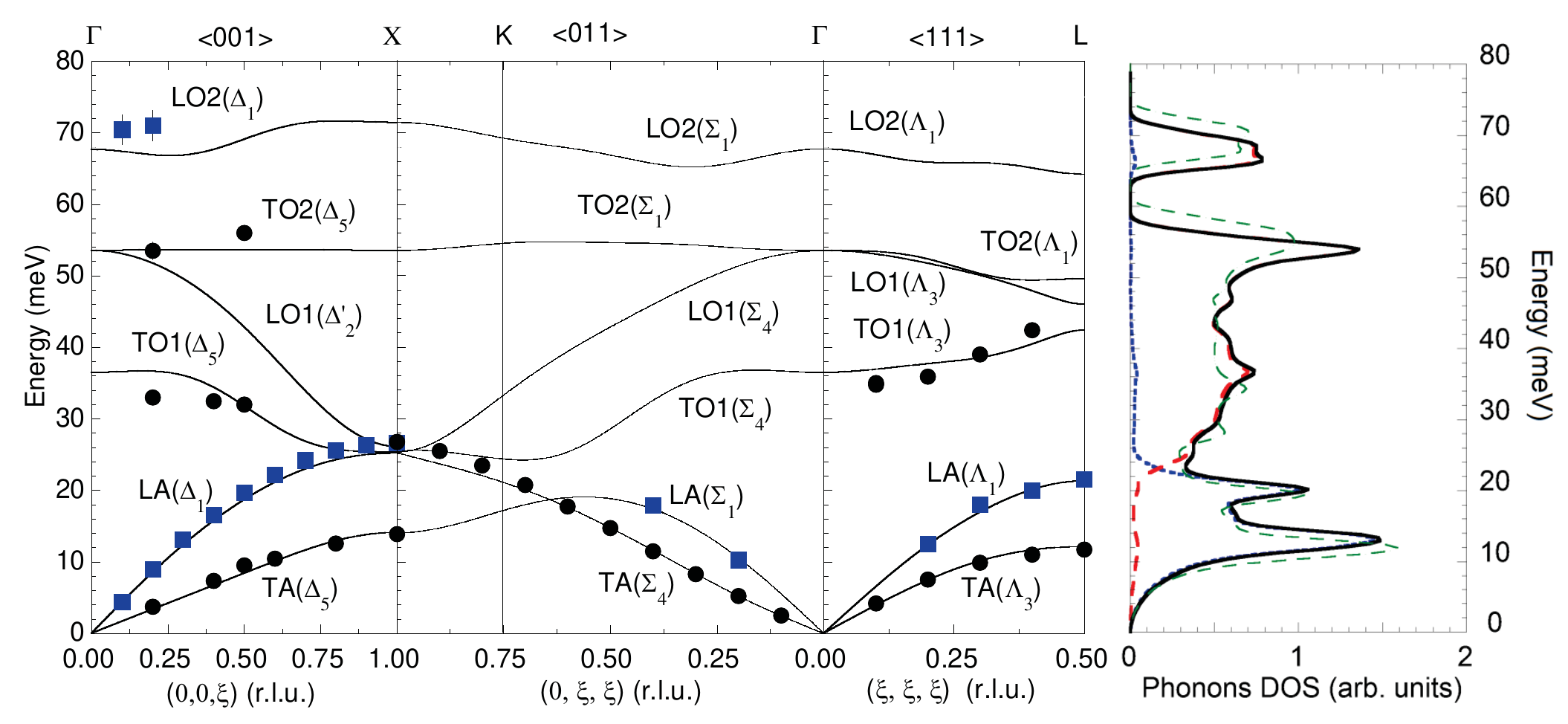}
\caption{(Color online) Phonon dispersion relations of NpO$_{2}$ measured at room temperature along three high-symmetry directions by IXS (symbols) and calculated (solid lines) by first-principles density functional theory in the GGA+$U$ approach ($U = 4$ eV,  $J = 0.6$ eV). $\xi$ is the phonon wavevector component in reciprocal lattice units (r.l.u.) of  $2\pi/a$, where $a$ is the lattice parameter. Transverse (longitudinal) modes are indicated by black circles (blue squares). The labels of the branches refer to the irreducible representations of the wavevector little group. The total phonon density of states (pDOS) is shown in the right panel by the solid black line. Partial pDOS for Np and O atoms are indicated by dotted blue line and dashed red line, respectively. The Np atom contribution to the pDOS dominates below $\sim$ 25 meV, whereas the higher energy contributions are dominated by the light O atoms vibrations.
The pDOS simulated for UO$_{2}$ is shown for comparison (thin dashed green line).
\label{dispersion}}
\end{figure*}

Figure \ref{dispersion} shows the phonon dispersion curves at 300 K calculated \textit{ab initio} together with data points measured along the three main symmetry directions $[001]$ ($\Delta$), $[110]$ ($\Sigma$), and $[111]$ ($\Lambda$), the experimental errors being smaller than the size of the symbols. The nine phonon branches for the NpO$_{2}$ cubic cell [symmorphic CaF$_{2}$-type \textit{fcc} structure with Space Group $Fm\bar{3}m$ and three atoms per unit cell, at $\bm{R}_{\rm O} = \pm(a/4)(1, 1, 1)$ and $\bm{R}_{\rm Np} = (0, 0, 0)$] contain double-degenerate transverse acoustic (TA) and transverse optic (TO1; TO2) modes, one
longitudinal acoustic (LA) mode, and two longitudinal optic (LO1; LO2) modes. The group theoretical notation in Fig.\ \ref{dispersion} refers to the irreducible representations of the little group of the phonons wavevector. The atom-projected phonon density of states (pDOS) is shown in the right-hand panel of Fig.\ \ref{dispersion}. At low energies the contribution from the Np atoms is larger than that from the O atoms, while at higher vibrational frequencies this behavior is reversed, as expected from the masses of the atomic components.

{\red The measured and \textit{ab initio} computed phonon dispersions are overall in good agreement with another. Only the LO1 branch could not be observed, which has a very small Np contribution.}
Compared with UO$_{2}$ {\red (Refs.\ \onlinecite{pang13} and \onlinecite{pang14})}, the main differences in the pDOS are a softening of the optical modes and an increase of the peak centered around 55 meV and corresponding to the TO2 modes, whereas the acoustic modes in NpO$_{2}$ are shifted to slightly higher frequencies.
%{\red this last part needs references}

\subsection{Main thermodynamic properties}

From the Gibbs free energy the main thermodynamic parameters such as the bulk modulus $B$, the thermal expansion coefficient $\alpha$, and the lattice isobaric heat capacity $C_p$ can be obtained
\begin{eqnarray}
B &=& V\left(\frac{\partial^2 F}{\partial V^2} \right)_{\! T} , \\
\alpha &=&\frac{1}{V}\left(\frac{\partial}{\partial T}\left(\frac{\partial G}{\partial P} \right)_{\! T} \right)_{\! p}  ,\\
C_p &=&-T\left(\frac{\partial^2 G}{\partial T^2}\right)_{\! p} .
\end{eqnarray}

\begin{figure}
\includegraphics[width=8cm]{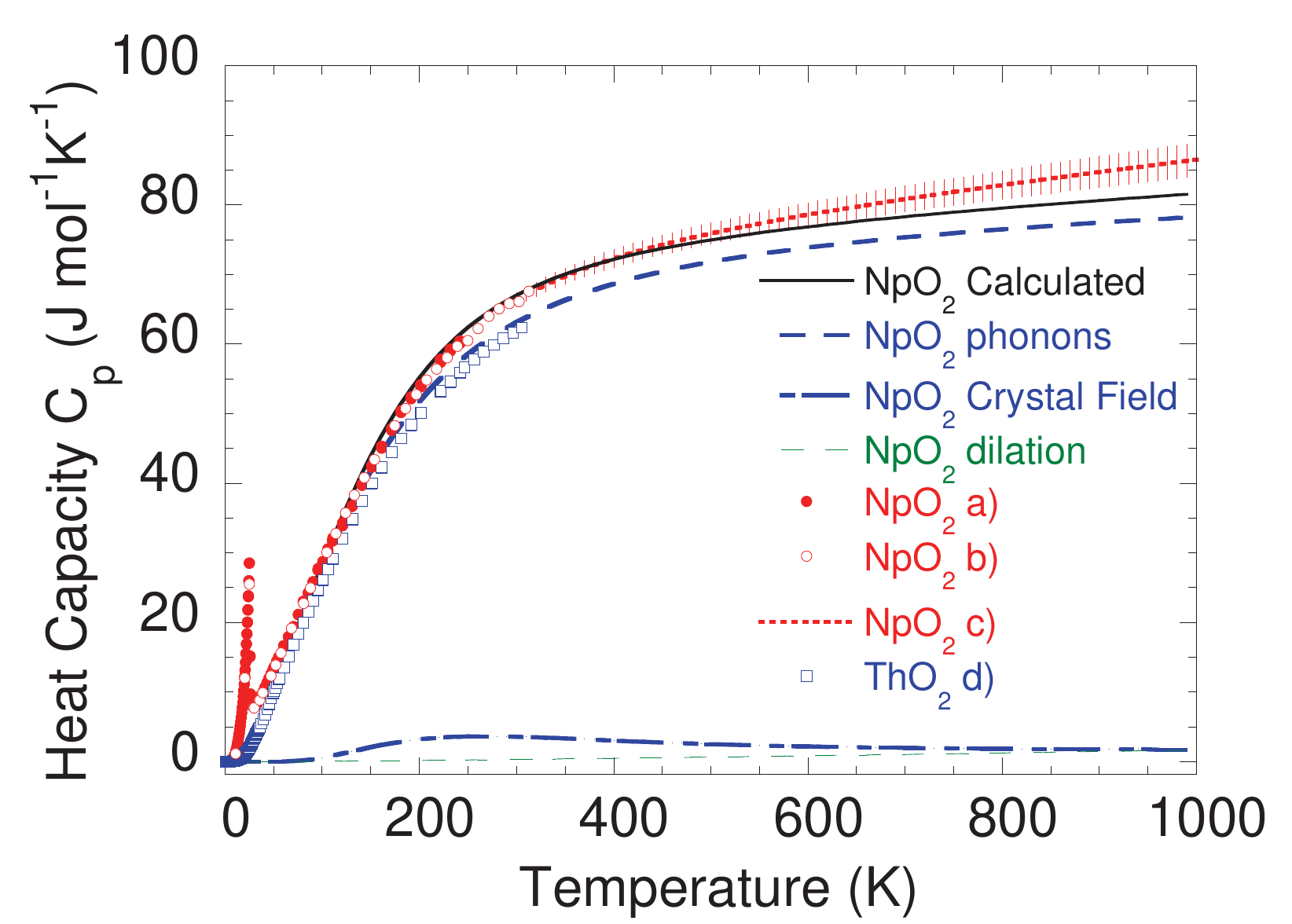}
\caption{(Color online) Constant-pressure heat capacity of NpO$_{2}$ (black solid line) obtained as the sum of the calculated vibrational contribution (blue dashed line), the dilation contribution calculated by the Gr\"{u}neisen relation (green dashed line), Ref.\
[\onlinecite{serizawa01}], and  the Schottky contribution due to crystal field (CF) excitations (blue dot-dashed line), calculated on the {\red basis} of the CF level scheme derived from inelastic neutron scattering experiments \cite{magnani05,fournier91,caciuffo92}. Experimental data are taken from Ref.\ [\onlinecite{magnani05a}] (a, red solid circles), Ref.\ [\onlinecite{westrum53}] (b, red open circles), and Ref.\  [\onlinecite{benes11}] (c, red dot line). The lambda-type anomaly at 25 K is  associated with the transition to the multipolar ordered phase of NpO$_{2}$. Experimental data for ThO$_{2}$ (d, blue open squares) are taken from Ref.\ [\onlinecite{osborne53}].  \label{Cp-temperature}}
\end{figure}

In Fig.\ \ref{Cp-temperature} we present the \textit{ab initio} calculated lattice heat capacity $C_p$ as a function of temperature, compared with available experimental data
%from different works
\cite{magnani05a,westrum53,benes11}. The calculated curve provides a very good {\red description} of the data measured for the empty $5f$ shell analog ThO$_{2}$ \cite{osborne53}. The agreement with the NpO$_{2}$ data
is also good for temperatures in the range between 50 and 200 K. At lower temperatures the phase transition to the multipolar ordered phase is responsible for the appearance of a large anomaly, which is not considered in the theory discussed in this work. At higher temperatures the Schottky contribution becomes relevant, as shown in Fig.\ \ref{Cp-temperature} by the blue dot-dashed line that corresponds to the crystal field (CF) energy levels scheme provided by inelastic neutron scattering experiments \cite{magnani05,fournier91,caciuffo92}. A dilation contribution to the specific heat curve must also be considered. This can be estimated using  the Gr\"{u}neisen relation, $C_{d} = \alpha \gamma T C_{p}$, where $\alpha$ is the thermal expansion coefficient, $T$ the temperature, and $\gamma$ is the Gr\"{u}neisen parameter \cite{serizawa01}. Our calculations provide a value of $\gamma \simeq$~2, almost temperature independent in the $T$ range between 300 and 1000 K. Using the experimental thermal expansion data reported in Ref.\ [\onlinecite{serizawa01}], we obtain the green dashed line in Fig.\ \ref{Cp-temperature}. The total heat capacity obtained by summing vibrational, Schottky, and dilation contributions is in good agreement with the experiments up to 1000 K.

\begin{figure}
\includegraphics[width=8cm]{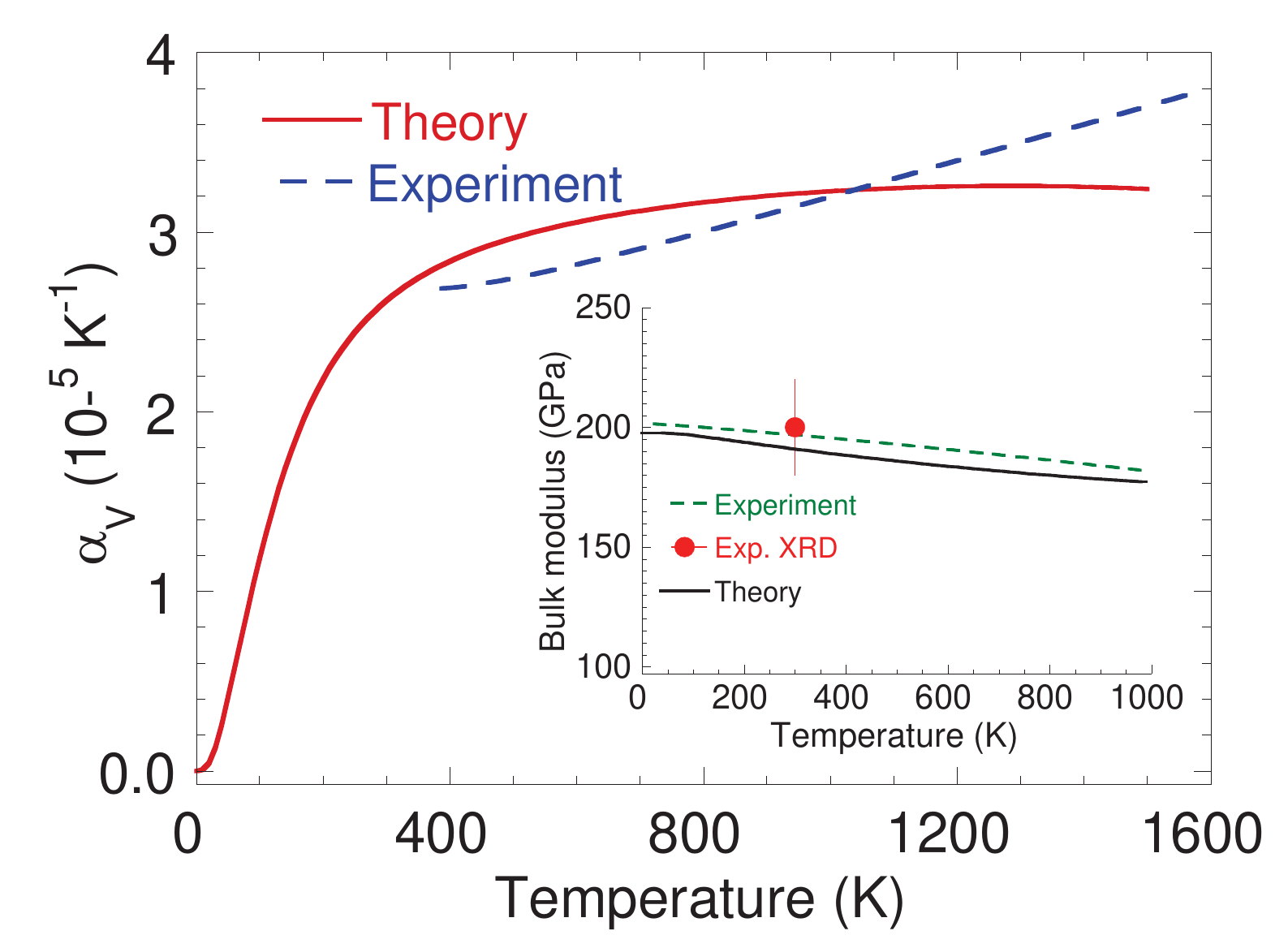}
\caption{(Color online) Calculated (red solid line) and experimental thermal expansion of NpO$_2$ (blue dashed line, from Ref.\ [\onlinecite{serizawa01}]). The inset shows the calculated (black solid line) bulk modulus of NpO$_2$ compared with the experimental value provided by high-pressure X-ray diffraction measurements \cite{benedict86} (red circle) and the analytical estimation given in Ref.\ [\onlinecite{sobolev09}] (green dashed line). \label{alpha}}
\end{figure}

For the thermal expansion coefficient we compare our \textit{ab initio} results with the available experimental data obtained by Serizawa \textit{et al.} \cite{serizawa01} in the temperature range of $298-1600$ K. The results, plotted in Fig.\ \ref{alpha}, show that for temperatures below 800 K the qualitative behavior between the experimental and simulated thermal expansion is the same, apart a slight overestimation of the theoretical results. However, above 800 K the trend for the simulated thermal expansion changes, becoming unexpectedly almost constant and unlike the experimental results which still show an almost linear increase. The reason for the discrepancy between experimental and simulated results has already been pointed out in the previous section, namely the validity of the QHA at high temperatures. Since the main thermodynamic properties, such as heat capacity, thermal expansion or melting point, of the actinide oxides (AnO$_2$ with An being Th, U, Np or Pu) are very similar, our results emphasize the limitation of QHA to reproduce the right expanded lattice for these compounds at high temperatures (for instance, the phonon dispersion calculated for UO$_{2}$ at $T=1200$ K by Pang and coworkers in Ref.\ [\onlinecite{pang13}] most probably suffers from being based on a wrongly expanded lattice). Yun \textit{et al.} also employed the QHA to compute the thermal expansion coefficient of UO$_2$ and obtained reasonable agreement with experiment up to 500 K \cite{yun12}. On the other hand, the influence of this limitation seems to be less relevant when computing the heat capacity and bulk modulus as shown above.

The calculated bulk modulus as a function of temperature is presented in the inset of Fig.\ \ref{alpha}. The computed value at 0 K of 197.7 GPa is in agreement with the only available experimental value ($B = 200$ GPa) \cite{benedict86} as well as with previous theory \cite{wang10}. The temperature-evolution of this quantity can be compared with values obtained by Sobolev \cite{sobolev09}, who proposed a set of analytical models for the calculation of the main thermophysical properties of the actinide dioxides based on a simplified phonon spectrum, the quasiharmonic approximation for the lattice vibrations, and the Klemens approach for the thermal conductivity. We find that the agreement in the whole plotted temperature range is very good.

\subsection{Thermal conductivity}

Next, we investigate the lattice thermal conductivity $\kappa$. This quantity is represented by a second-rank tensor that in cubic crystals, as NpO$_2$, becomes isotropic and can therefore be described by a single value. As a consequence, the velocity and transport lifetime components are in the heat transport direction, parallel to a small applied temperature gradient.
It has been suggested by Gofryk \textit{et al.} \cite{gofryk14} that the presence of temperature gradients interacting with the electronic moments on UO$_{2}$ breaks the cubic symmetry,
inducing thus an anisotropic thermal conductivity. A similar phenomenon could occur in NpO$_{2}$. However, in our simulations we have not considered the temperature gradients, and therefore our computational results corroborate the crystallographic symmetry assumptions in the whole range of considered temperatures.

\begin{figure}
\includegraphics[width=8cm]{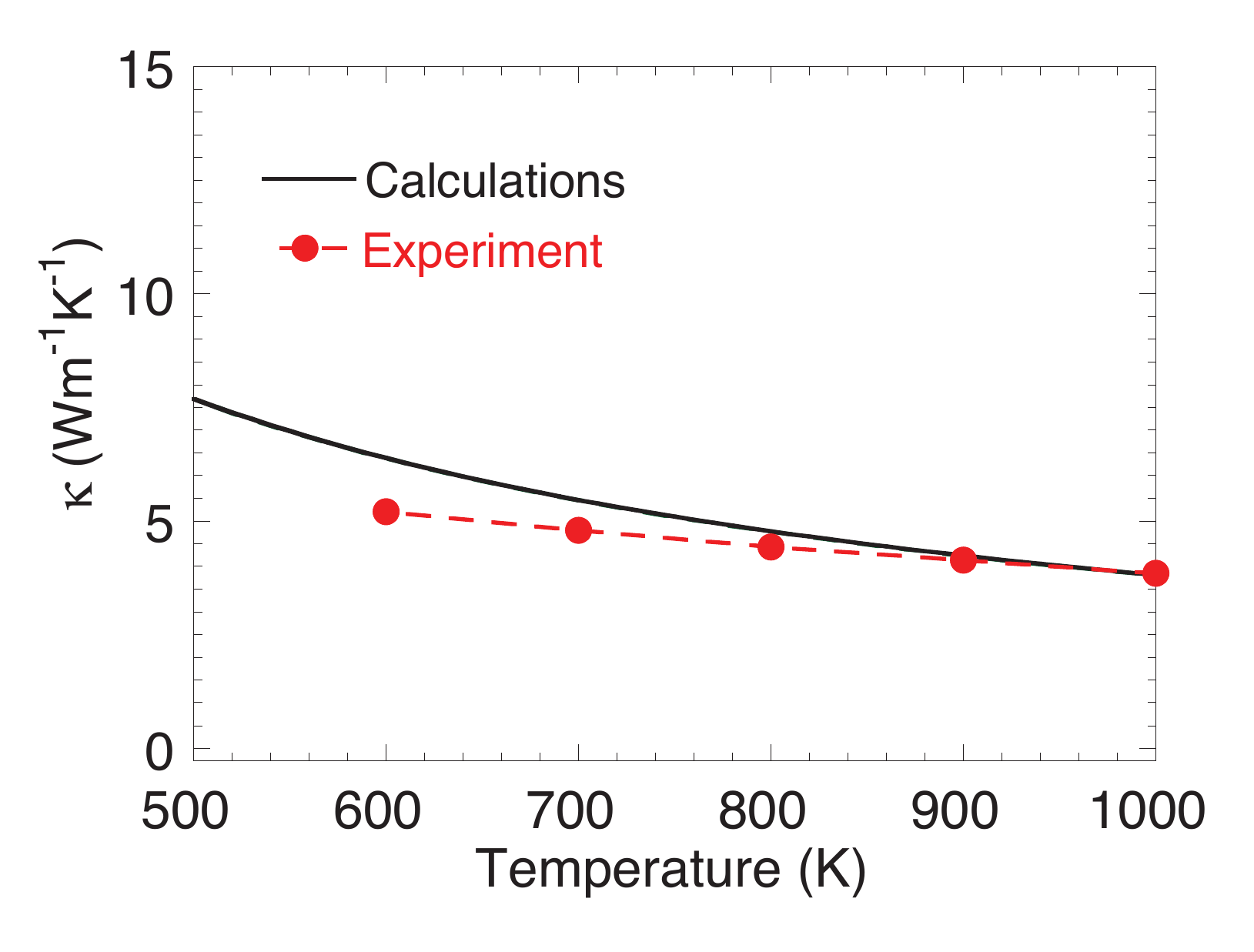}
\caption{(Color online) Calculated temperature dependence of the thermal conductivity of the $fcc$ NpO$_2$ crystal using the full solution of the Boltzmann equation (black line) from first-principles anharmonic lattice dynamics calculations. Simulations based on the RTA approximation provide undistinguishable results in the shown temperature range. The experimental thermal conductivity curve given in Ref.\ [\onlinecite{minato09}] is shown by the red circles with connecting line. \label{kappa}}
\end{figure}

In Fig.\ \ref{kappa} we show the temperature dependence of $\kappa$ computed from first-principles anharmonic lattice dynamics calculations with both the RTA and the full solution of the Boltzmann equation for temperatures ranging from 500 to 1000 K. {\red Note that} the two approaches provide undistinguishable results. The comparison of the calculated curve with the one derived from the experimental values for
thermal diffusivity, bulk density and specific heat \cite{nishi08,minato09} shows a good agreement and supports the validity of the RTA approach in the whole range of considered temperatures. Here it is important to emphasize that for a correct description of the thermal conductivity at lower temperatures it is necessary to take into consideration the effect of crystalline grain boundaries, point defects and isotopic scattering. Due to the lack of experimental results at low temperature and the fact that those contributions to the phonon-phonon scattering are sample dependent, we have not addressed the low temperature regime. However, our results can be easily extended to include this contribution by evaluating the phonon lifetime in Eq.\ (\ref{eq-RTA}) as
\begin{equation}
\frac{1}{\tau}=\frac{1}{\tau_{gb}}+\frac{1}{\tau_{pd}}+\frac{1}{\tau_{is}}+\frac{1}{\tau_{ph-ph}},
\end{equation}
where $\tau_{gb}$, $\tau_{pd}$, $\tau_{Is}$, and $\tau_{ph-ph}$ are respectively the phonon lifetimes due to phonon scattering by grain boundaries, point defects, isotopic scattering and anharmonicities. Furthermore, it is important to notice that the temperature dependence of the thermal conductivity is calculated for the structure at $T=0$ K. The thermal expansion, which will contribute to slightly decrease the {\red computed} thermal conductivity, is not taken into account. The negligible contribution from this effect is evidenced by the good agreement obtained between the theoretical and experimental results.

\begin{figure*}
\includegraphics[width=0.8\linewidth]{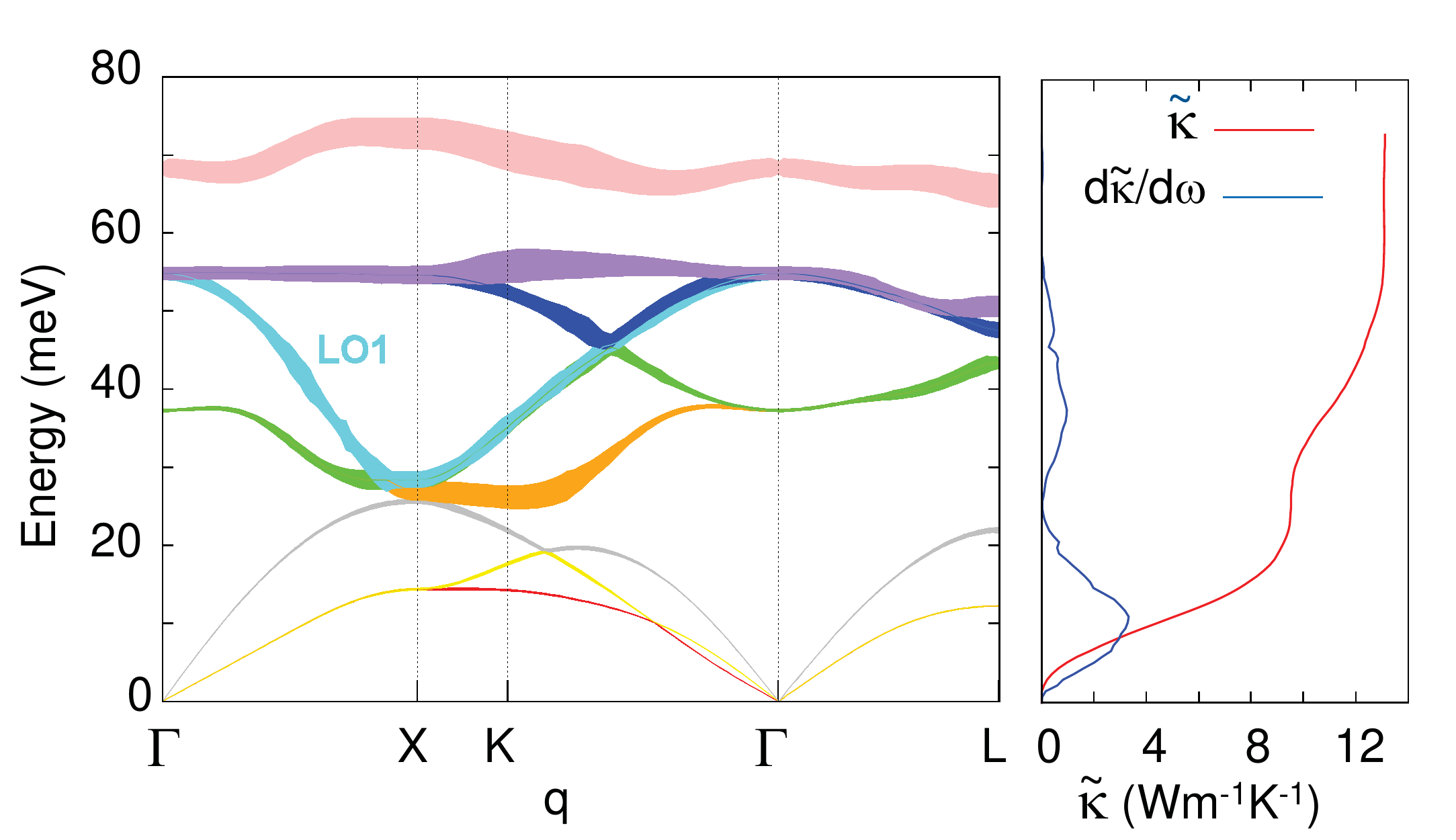}
\caption{(Color online) Phonon linewidth distribution calculated along the high-symmetry lines for \textit{fcc} NpO$_2$  at 300 K (left panel) and accumulated phonon thermal conductivity $\tilde{\kappa}$ as a function of the phonon energy (right panel). Different colors in the linewidth distribution indicate different phonon branches, {\red whilst the $q$-dependent linewidth is depicted by the width of the branch.} The accumulated phonon thermal conductivity is {\red shown} {\cyan (right panel)} by the red line whilst its derivative with respect to the energy is depicted by the blue line.
\label{acc-kappa}}
\end{figure*}

\begin{figure*}
\includegraphics[width=14cm]{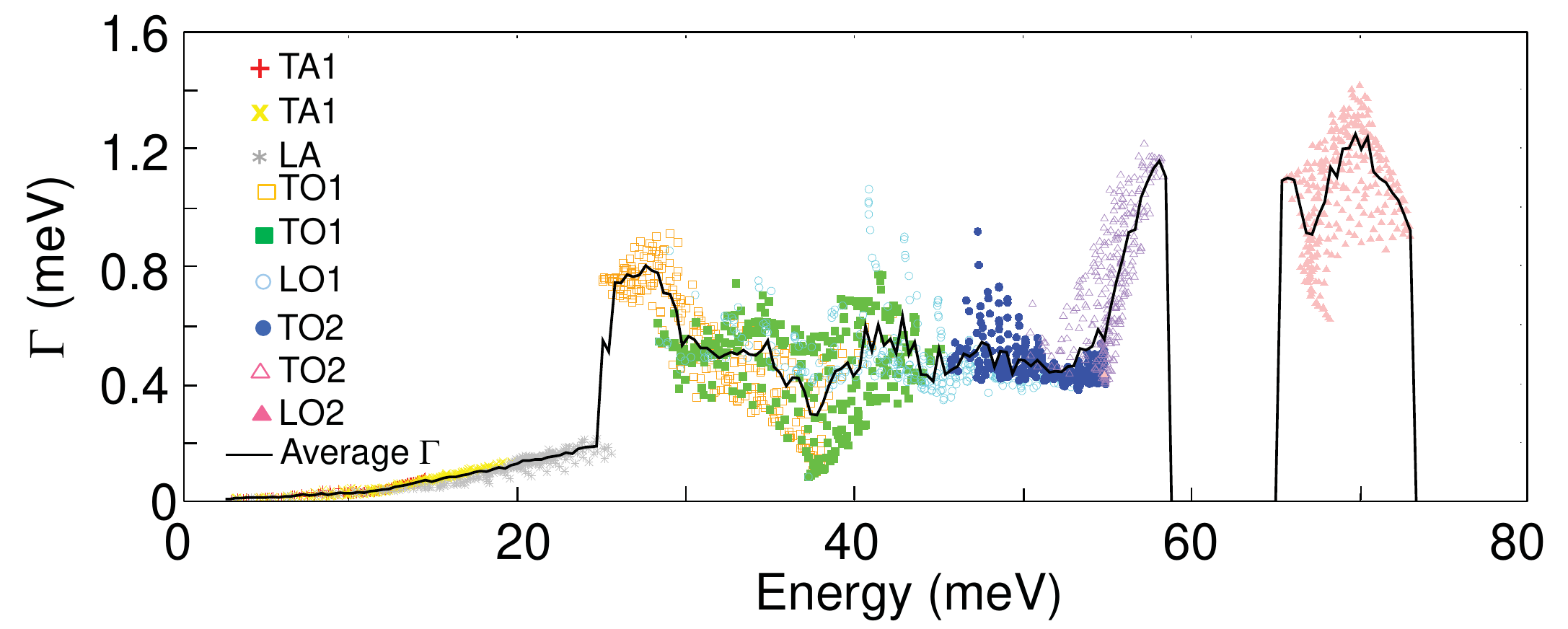}
\caption{(Color online) Variation of the half of the phonon linewidth for NpO$_2$ with respect to the phonon frequency. Different phonon branches are indicated with different colors, while the average phonon linewidth is represented by a black line.\label{linewidth}}
\end{figure*}

We now evaluate the contribution of each phonon frequency to the total lattice thermal conductivity at 300 K, by integrating $\kappa$ calculated within the RTA methodology for all the frequencies spanned by the phonons. The accumulated thermal conductivity, $\tilde{\kappa}$, plotted in Fig.\ \ref{acc-kappa}, shows the increase of $\kappa$ with increasing frequency. To better illustrate the variation of this quantity we also show its derivative with respect to the energy, $d\tilde{\kappa}/d\omega$. Remarkably, our results show that phonons with energy below 6 THz ($\sim$ 25 meV) only contribute to {\red 73}\% of the total thermal conductivity, while the remaining {\red 27}\% is due to phonons with a higher energy. This is a striking result, which emphasizes the importance of high-energy optical phonons in the heat transport. To better illustrate this fact, we show the calculated phonon linewidths in Fig.\ \ref{acc-kappa} along high-symmetry lines in the BZ. The smaller linewidth along with the larger group velocities (steeper bands) of the acoustic phonons compared with those of the optical phonons translate into a larger contribution to the total thermal conductivity, {\red cf.\ Eq.\ (\ref{eq-RTA})}. However, the still small linewidth of the TO1 phonons (see {\green Fig.}\ \ref{linewidth} {\red below}) and their high velocities give that TO1 and LO1 modes transport a relatively large amount of heat. In contrast, TO2 and LO2 branches have a very weak contribution to the total thermal conductivity, mainly due to the flat character of the dispersion bands
{\red i.e., small velocity}. Our conclusions are in agreement with previous theoretical and experimental results for UO$_2$ where it was shown that the largest amount of heat is transported by the LO1 phonon mode \cite{pang13,pang14}. {\red Unfortunately, this mode contains almost no contribution from the metal atom, hence it is very hard to observe with IXS, and we did not manage to establish its energy dispersion.}

At higher temperatures a decrease of the thermal conductivity is expected due to the broadening of the linewidths with temperature. In Fig.\ \ref{linewidth} we show the {\red calculated} changes of the phonon linewidth with the phonon frequency at $T = 300$ K. As expected, the acoustic phonons present the lowest linewidths while the LO2 modes show the largest. However, in agreement with the results for the thermal conductivity, we also find that the linewidths of {\red the} TO1 phonons are of the same order of those of {\red the} acoustic phonons in the range of 8.5 to 9.5 THz ($\sim 35-39$ meV).

\begin{figure*}
\includegraphics[width=14cm]{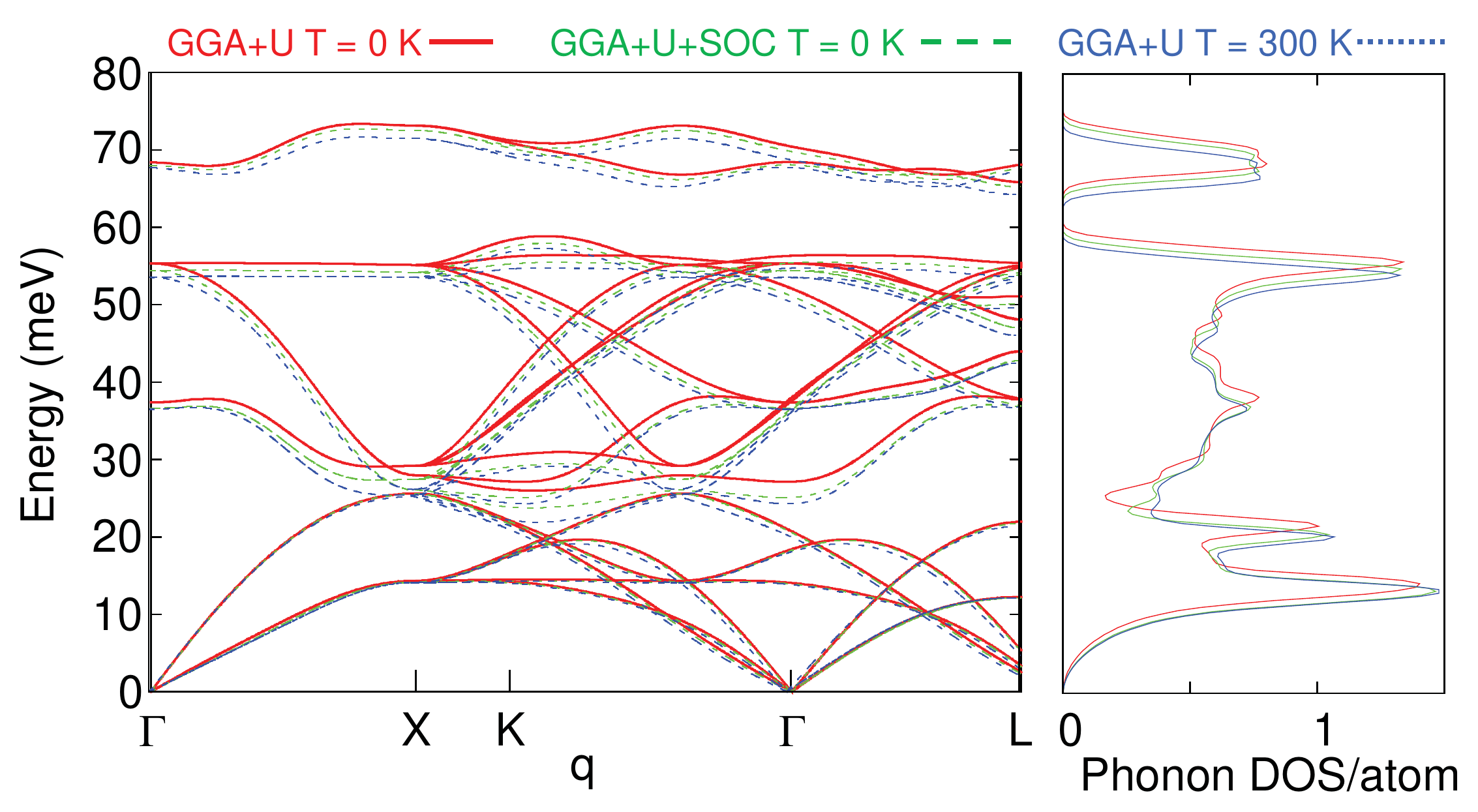}
\caption{(Color online) GGA+$U$ at 0 K (red line), GGA+$U$+SOC at 0 K (green dashed line) and GGA+$U$ at 300 K (blue dashed line) calculated phonon dispersions (left panel) and corresponding phonon density of states (right panel) of NpO$_2$. {\red Note that the spin-orbit coupling has a relatively small influence on the \textit{ab initio} computed phonon dispersions, comparable to that of an increased temperature.}
%The $\bm{q}$-point labels in the left panel are those for the standard high-symmetry positions of the fcc Brillouin zone
\label{fig-SOC}}
\end{figure*}

{\red Lastly,} to evaluate the effect of {\red the} spin-orbit coupling on the phonon properties, {\red we have carried out} a calculation of the phonon dispersions including the SOC {\red in the electronic structure.} The results, shown in the left panel of Fig.\ \ref{fig-SOC}, are compared with the phonon dispersions {\red computed} without SOC along the {\red high-}symmetry lines at $T=0$  and 300 K. {\red Accounting for the SOC leads on}
 average {\red to a phonon} softening {\red comparable} to an increase of the temperature of the system. This can be more easily appreciated in the right panel of Fig.\ \ref{fig-SOC}, where the phonon density of states is shown. The computed densities of states highlight the relatively small importance of the SOC on the thermal conductivity of the system that, as shown above, has a negligible dependence on the thermal expansion.

\begin{figure*}
\includegraphics[width=14cm]{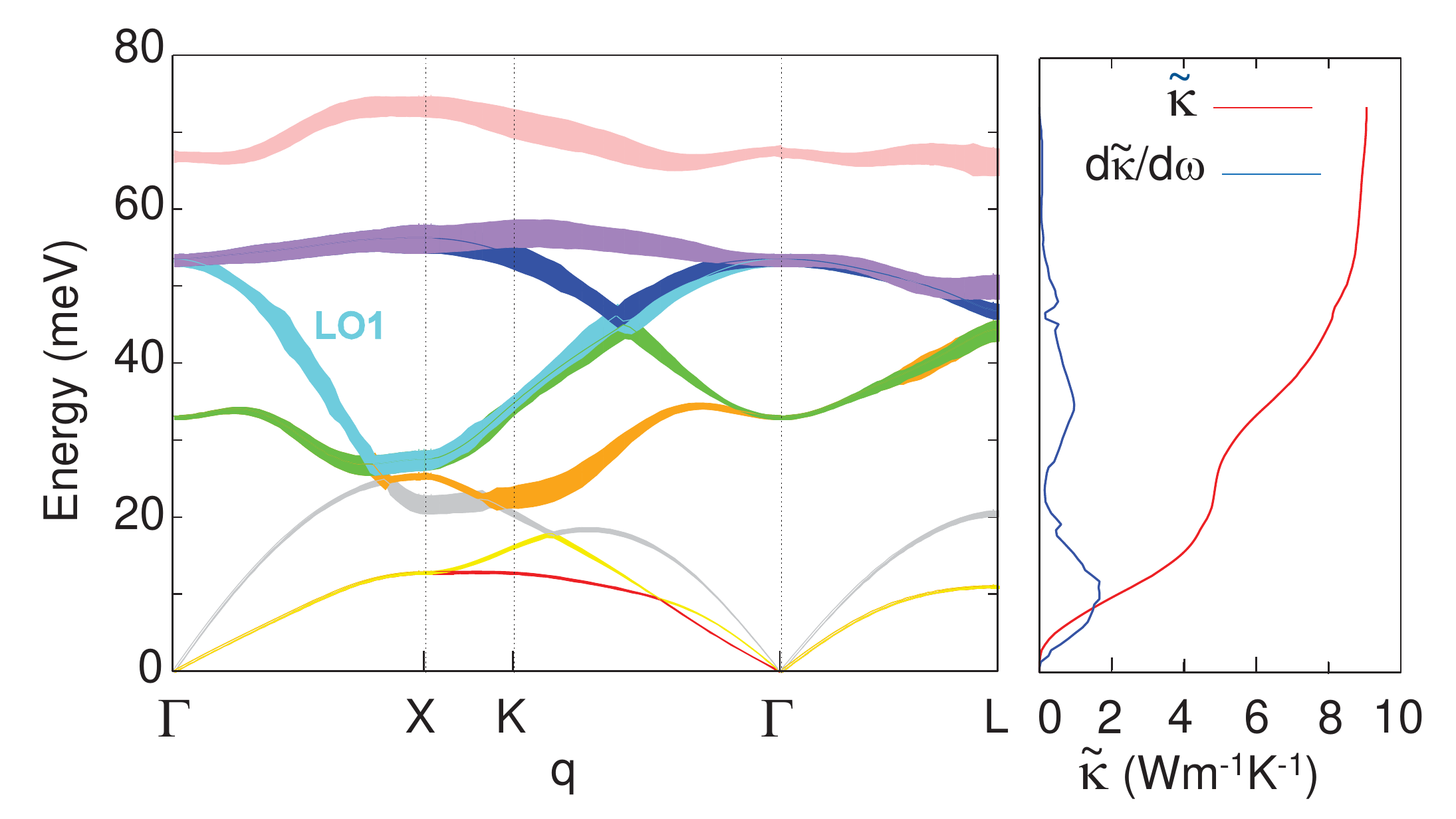}
\caption{(Color online) Phonon linewidth distribution calculated along the high-symmetry lines for \textit{fcc} UO$_2$  at 300 K (left panel) and accumulated phonon thermal conductivity $\tilde{\kappa}$ as a function of the phonon energy (right panel). Different colors in the linewidth distribution indicate different phonon branches, and the $q$-dependent linewidth is depicted by the width of the branch. The accumulated phonon thermal conductivity is shown {\cyan (right panel)} by the red line whilst its derivative with respect to the energy is depicted by the blue line.}
\label{acc-kappa-UO2}
\end{figure*}

\section{Discussion and Conclusions}

%{\red still to be done - comparison with UO2, see G.L. remarks; (1) thermal cond., (2) bulk modulus and (3) phonon DOS, and further points}

{\red
The phonons and thermal behavior of UO$_2$ have been extensively investigated in the past, both experimentally and theoretically \cite{dolling65,pang13,pang14,gofryk14,yin08,sanati11,yun12}. This makes  UO$_2$ an ideal material for comparison with our data obtained for NpO$_2$. An issue that recently has emerged is the origin of the lattice thermal conductivity of UO$_2$, specifically, which phonon modes are carriers of the thermal conductivity. Dynamical mean field theory (DMFT) calculations predicted that only LA modes contribute to the heat transport \cite{yin08}. More recent measurements and \textit{ab initio} calculations on the basis of the GGA+$U$ method came to a different conclusion, namely, that LO phonons transport a surprisingly large fraction of 30\% of the total heat \cite{pang13,pang14}. We observe that our calculations for NpO$_2$ are in good accordance with the latter work on UO$_2$, as we find that 27\% of the total heat in NpO$_2$ is transported by the optical modes. The low contribution of the LO phonons to the total heat transport in UO$_2$, as predicted by DMFT, could be related to the predicted {\blue steeply} dispersive LO phonons, giving low group velocities that are $2-3$ lower than obtained in measurements. The DMFT predicted thermal conductivity at 1000~K was also a factor of two lower than the experimental value of 3.9 Wm$^{-1}$K$^{-1}$
(Refs.\ \onlinecite{hyland83} and \onlinecite{fink00}). For comparison, the thermal conductivity of NpO$_2$, computed here with the GGA+$U$ method is 4.2 Wm$^{-1}$K$^{-1}$ at 1000~K, i.e., quite close to the value known for UO$_2$, and in good agreement with the measured value \cite{nishi08} of 4.1 Wm$^{-1}$K$^{-1}$ for NpO$_2$. NpO$_{2}$ has a smaller thermal conductivity than UO$_{2}$, at least in the temperature range 600 to 1000 K for which experimental values are available (at 600 K, $\kappa$(UO$_{2}$) is about 31\% larger than $\kappa$(NpO$_{2}$) \cite{minato09}. This is accounted for by our simulation, which however overestimates the experimental value for NpO$_{2}$ by $\sim$20\% at 600 K.

{\cyan For further comparison with UO$_2$ we provide in Appendix A (Fig.\ \ref{acc-kappa-UO2}) the \textit{ab initio} calculated phonon dispersions and phonon linewidths of UO$_2$ at 300~K.}
Compared with UO$_{2}$, the main differences in the calculated phonon DOS are a softening of the optical modes and an increase of the peak centered around 55 meV (TO2 modes), whereas the acoustic modes in NpO$_{2}$ are shifted to slightly higher frequencies {\cyan (cf.\ Fig.\ \ref{dispersion})}. The calculated value  of the bulk modulus in NpO$_2$ ($B = 197.7$ GPa, at 0 K) is in good agreement with the experimental value ($B = 200$ GPa) \cite{benedict86} and slightly smaller than the one determined by high-pressure X-ray diffraction for UO$_{2}$ [$B$(UO$_{2}$) $= 207\pm2$ GPa] \cite{idiri04}.
{\cyan The computed accumulated phonon thermal conductivity of UO$_2$ is also shown in the Appendix. Consistent with our observations for NpO$_2$ we find that the optical phonon branches do contribute significantly to the lattice thermal conductivity.
}

The good agreement of the measured  and \textit{ab initio} calculated phonon dispersions indicates that the GGA+$U$ method provides a good description of the crystal lattice vibrations of NpO$_2$ and thus permits to perform \textit{ab initio} studies of its thermodynamic properties. We find that  the calculated heat capacity and bulk modulus are in good agreement with available measured quantities. The calculated thermal expansion however agrees with experiment only up to 1000~K. The failure in the description of the thermal expansion at higher temperatures indicates the limitation of the quasiharmonic approximation to study actinide dioxides beyond $\sim$ 1000 K. {\blue The phonon linewidths and} the thermal conductivity of NpO$_2$ have been for the first time evaluated using first-principles anharmonic lattice dynamics simulations. We obtain an excellent agreement with the scarcely available experimental values {\blue for the thermal conductivity}. We {\blue have} established that optical phonons contribute significantly to the heat transport (by about 27\%), mainly due to their large velocities and short lifetimes, which, for TO1 phonons are comparable to acoustic phonons. Taking the spin-orbit interaction into account in the first-principles calculations leads to a small softening of the phonon modes, an effect which is comparable to an increase of the lattice temperature, and which results in a negligible influence on the phononic thermal properties.

Our first-principles simulations and measurements of the phononic properties of NpO$_2$ are relevant for the modeling  of nuclear fuel materials, for which high thermal conductivities are desirable. Our results demonstrate that the density functional theory at the level of GGA+$U$ approach, although usually not sufficient to describe low-energy scale interactions as the Kondo screening, works very well to simulate quantities associated with larger energy scales, such as lattice vibrations and structural, mechanical and thermodynamical properties.

{\green{Finally, as only very small crystals are required, the methods described in this article can be easily extended to PuO$_{2}$ and AmO$_{2}$, for which no experimental data on the dispersion of phonon branches are available.}}

\acknowledgments
{We thank G.\ Pagliosa for technical support during the encapsulation of the NpO$_{2}$ single crystal. We are grateful to P.\ Colomp and the ESRF radioprotection service for their cooperation during the experiment. We acknowledge financial support from the Swedish Research Council (VR) and the Knut and Alice Wallenberg Foundation, as well as support from the Swedish National Infrastructure for Computing (SNIC).

%\clearpage
%\noindent%

\appendix
\section{Comparison to UO$_2$}
\label{appendix}

The lattice vibrations contribution to the thermal properties of UO$_2$ (a key element for the safety assessment of the operation of nuclear power plants) have been extensively investigated \cite{dolling65,pang13,pang14,gofryk14}.
To perform a  comparison between UO$_2$ and NpO$_2$ we have applied the computational methodology described above to simulate the phonon dispersion and phonon lifetimes of UO$_2$. Figure \ref{acc-kappa-UO2}
shows the dispersion relations along the high-symmetry direction of the UO$_2$ {\cyan \textit{fcc}} lattice, together with the accumulated phonon thermal conductivity. The {\cyan \textit{ab initio}} calculated value of the latter quantity ($\tilde{\kappa}_{calc}$ = 9.06 Wm$^{-1}$K$^{-1}$) compares well with the experimental value $\tilde{\kappa}_{INS}$ = 8.4$\pm$1.5 Wm$^{-1}$K$^{-1}$ obtained at 295 K by Pang \textit{et al.} from inelastic neutron scattering measurements \cite{pang13} and with the macroscopic room temperature thermal conductivity reported by Hyland ($\tilde{\kappa}_{mac}$ = 9.7 Wm$^{-1}$K$^{-1}$) \cite{hyland83}.
{\cyan Note that for UO$_2$ the optical phonon branches contribute significantly to the total lattice thermal conductivity.}

%\clearpage
\bibliography{NpO2phonons}

\begin{thebibliography}{57}
\expandafter\ifx\csname natexlab\endcsname\relax\def\natexlab#1{#1}\fi
\expandafter\ifx\csname bibnamefont\endcsname\relax
  \def\bibnamefont#1{#1}\fi
\expandafter\ifx\csname bibfnamefont\endcsname\relax
  \def\bibfnamefont#1{#1}\fi
\expandafter\ifx\csname citenamefont\endcsname\relax
  \def\citenamefont#1{#1}\fi
\expandafter\ifx\csname url\endcsname\relax
  \def\url#1{\texttt{#1}}\fi
\expandafter\ifx\csname urlprefix\endcsname\relax\def\urlprefix{URL }\fi
\providecommand{\bibinfo}[2]{#2}
\providecommand{\eprint}[2][]{\url{#2}}

\bibitem[{\citenamefont{Brockhouse and Stewart}(1958)}]{brockhouse58}
\bibinfo{author}{\bibfnamefont{B.~N.} \bibnamefont{Brockhouse}}
  \bibnamefont{and} \bibinfo{author}{\bibfnamefont{A.~T.}
  \bibnamefont{Stewart}}, \bibinfo{journal}{Rev. Mod. Phys.}
  \textbf{\bibinfo{volume}{30}}, \bibinfo{pages}{236} (\bibinfo{year}{1958}).

\bibitem[{\citenamefont{Dolling et~al.}(1965)\citenamefont{Dolling, Cowley, and
  Woods}}]{dolling65}
\bibinfo{author}{\bibfnamefont{G.}~\bibnamefont{Dolling}},
  \bibinfo{author}{\bibfnamefont{R.~A.} \bibnamefont{Cowley}},
  \bibnamefont{and} \bibinfo{author}{\bibfnamefont{A.~D.~B.}
  \bibnamefont{Woods}}, \bibinfo{journal}{Can. J. Phys.}
  \textbf{\bibinfo{volume}{43}}, \bibinfo{pages}{1397} (\bibinfo{year}{1965}).

\bibitem[{\citenamefont{Pang et~al.}(2013)\citenamefont{Pang, Buyers,
  Chernatynskiy, Lumsden, Larson, and Phillpot}}]{pang13}
\bibinfo{author}{\bibfnamefont{J.~W.~L.} \bibnamefont{Pang}},
  \bibinfo{author}{\bibfnamefont{W.~J.~L.} \bibnamefont{Buyers}},
  \bibinfo{author}{\bibfnamefont{A.}~\bibnamefont{Chernatynskiy}},
  \bibinfo{author}{\bibfnamefont{M.~D.} \bibnamefont{Lumsden}},
  \bibinfo{author}{\bibfnamefont{B.~C.} \bibnamefont{Larson}},
  \bibnamefont{and} \bibinfo{author}{\bibfnamefont{S.~R.}
  \bibnamefont{Phillpot}}, \bibinfo{journal}{Phys. Rev. Lett.}
  \textbf{\bibinfo{volume}{110}}, \bibinfo{pages}{157401}
  (\bibinfo{year}{2013}).

\bibitem[{\citenamefont{Pang et~al.}(2014)\citenamefont{Pang, Chernatynskiy,
  Larson, Buyers, Abernathy, McClellan, and Phillpot}}]{pang14}
\bibinfo{author}{\bibfnamefont{J.~W.~L.} \bibnamefont{Pang}},
  \bibinfo{author}{\bibfnamefont{A.}~\bibnamefont{Chernatynskiy}},
  \bibinfo{author}{\bibfnamefont{B.~C.} \bibnamefont{Larson}},
  \bibinfo{author}{\bibfnamefont{W.~J.~L.} \bibnamefont{Buyers}},
  \bibinfo{author}{\bibfnamefont{D.~L.} \bibnamefont{Abernathy}},
  \bibinfo{author}{\bibfnamefont{K.~J.} \bibnamefont{McClellan}},
  \bibnamefont{and} \bibinfo{author}{\bibfnamefont{S.~R.}
  \bibnamefont{Phillpot}}, \bibinfo{journal}{Phys. Rev. B}
  \textbf{\bibinfo{volume}{89}}, \bibinfo{pages}{115132}
  (\bibinfo{year}{2014}).

\bibitem[{\citenamefont{Magnani
  et~al.}(2005{\natexlab{a}})\citenamefont{Magnani, Santini, Amoretti, and
  Caciuffo}}]{magnani05}
\bibinfo{author}{\bibfnamefont{N.}~\bibnamefont{Magnani}},
  \bibinfo{author}{\bibfnamefont{P.}~\bibnamefont{Santini}},
  \bibinfo{author}{\bibfnamefont{G.}~\bibnamefont{Amoretti}}, \bibnamefont{and}
  \bibinfo{author}{\bibfnamefont{R.}~\bibnamefont{Caciuffo}},
  \bibinfo{journal}{Phys. Rev. B} \textbf{\bibinfo{volume}{71}},
  \bibinfo{pages}{054405} (\bibinfo{year}{2005}{\natexlab{a}}).

\bibitem[{\citenamefont{Caciuffo et~al.}(1999)\citenamefont{Caciuffo, Amoretti,
  Santini, Lander, Kulda, and de~V.~Du~Plessis}}]{caciuffo99}
\bibinfo{author}{\bibfnamefont{R.}~\bibnamefont{Caciuffo}},
  \bibinfo{author}{\bibfnamefont{G.}~\bibnamefont{Amoretti}},
  \bibinfo{author}{\bibfnamefont{P.}~\bibnamefont{Santini}},
  \bibinfo{author}{\bibfnamefont{G.~H.} \bibnamefont{Lander}},
  \bibinfo{author}{\bibfnamefont{J.}~\bibnamefont{Kulda}}, \bibnamefont{and}
  \bibinfo{author}{\bibfnamefont{P.}~\bibnamefont{de~V.~Du~Plessis}},
  \bibinfo{journal}{Phys. Rev. B} \textbf{\bibinfo{volume}{59}},
  \bibinfo{pages}{13892} (\bibinfo{year}{1999}).

\bibitem[{\citenamefont{Carretta et~al.}(2010)\citenamefont{Carretta, Santini,
  Caciuffo, and Amoretti}}]{carretta10}
\bibinfo{author}{\bibfnamefont{S.}~\bibnamefont{Carretta}},
  \bibinfo{author}{\bibfnamefont{P.}~\bibnamefont{Santini}},
  \bibinfo{author}{\bibfnamefont{R.}~\bibnamefont{Caciuffo}}, \bibnamefont{and}
  \bibinfo{author}{\bibfnamefont{G.}~\bibnamefont{Amoretti}},
  \bibinfo{journal}{Phys. Rev. Lett.} \textbf{\bibinfo{volume}{105}},
  \bibinfo{pages}{167201} (\bibinfo{year}{2010}).

\bibitem[{\citenamefont{Caciuffo et~al.}(2011)\citenamefont{Caciuffo, Santini,
  Carretta, Amoretti, Hiess, Magnani, Regnault, and Lander}}]{caciuffo11}
\bibinfo{author}{\bibfnamefont{R.}~\bibnamefont{Caciuffo}},
  \bibinfo{author}{\bibfnamefont{P.}~\bibnamefont{Santini}},
  \bibinfo{author}{\bibfnamefont{S.}~\bibnamefont{Carretta}},
  \bibinfo{author}{\bibfnamefont{G.}~\bibnamefont{Amoretti}},
  \bibinfo{author}{\bibfnamefont{A.}~\bibnamefont{Hiess}},
  \bibinfo{author}{\bibfnamefont{N.}~\bibnamefont{Magnani}},
  \bibinfo{author}{\bibfnamefont{L.-P.} \bibnamefont{Regnault}},
  \bibnamefont{and} \bibinfo{author}{\bibfnamefont{G.~H.}
  \bibnamefont{Lander}}, \bibinfo{journal}{Phys. Rev. B}
  \textbf{\bibinfo{volume}{84}}, \bibinfo{pages}{104409}
  (\bibinfo{year}{2011}).

\bibitem[{\citenamefont{Amoretti et~al.}(1992)\citenamefont{Amoretti, Blaise,
  Caciuffo, Di~Cola, Fournier, Hutchings, Lander, Osborn, Severing, and
  Taylor}}]{amoretti92}
\bibinfo{author}{\bibfnamefont{G.}~\bibnamefont{Amoretti}},
  \bibinfo{author}{\bibfnamefont{A.}~\bibnamefont{Blaise}},
  \bibinfo{author}{\bibfnamefont{R.}~\bibnamefont{Caciuffo}},
  \bibinfo{author}{\bibfnamefont{D.}~\bibnamefont{Di~Cola}},
  \bibinfo{author}{\bibfnamefont{J.~M.} \bibnamefont{Fournier}},
  \bibinfo{author}{\bibfnamefont{M.~T.} \bibnamefont{Hutchings}},
  \bibinfo{author}{\bibfnamefont{G.~H.} \bibnamefont{Lander}},
  \bibinfo{author}{\bibfnamefont{R.}~\bibnamefont{Osborn}},
  \bibinfo{author}{\bibfnamefont{A.}~\bibnamefont{Severing}}, \bibnamefont{and}
  \bibinfo{author}{\bibfnamefont{A.~D.} \bibnamefont{Taylor}},
  \bibinfo{journal}{J. Phys. Condens. Matter} \textbf{\bibinfo{volume}{4}},
  \bibinfo{pages}{3459} (\bibinfo{year}{1992}).

\bibitem[{\citenamefont{Caciuffo et~al.}(1987)\citenamefont{Caciuffo, Lander,
  Spirlet, Fournier, and Kuhs}}]{caciuffo87}
\bibinfo{author}{\bibfnamefont{R.}~\bibnamefont{Caciuffo}},
  \bibinfo{author}{\bibfnamefont{G.~H.} \bibnamefont{Lander}},
  \bibinfo{author}{\bibfnamefont{J.~C.} \bibnamefont{Spirlet}},
  \bibinfo{author}{\bibfnamefont{J.~M.} \bibnamefont{Fournier}},
  \bibnamefont{and} \bibinfo{author}{\bibfnamefont{W.~F.} \bibnamefont{Kuhs}},
  \bibinfo{journal}{Solid State Commun.} \textbf{\bibinfo{volume}{64}},
  \bibinfo{pages}{149 } (\bibinfo{year}{1987}).

\bibitem[{\citenamefont{Santini et~al.}(2009)\citenamefont{Santini, Carretta,
  Amoretti, Caciuffo, Magnani, and Lander}}]{santini09}
\bibinfo{author}{\bibfnamefont{P.}~\bibnamefont{Santini}},
  \bibinfo{author}{\bibfnamefont{S.}~\bibnamefont{Carretta}},
  \bibinfo{author}{\bibfnamefont{G.}~\bibnamefont{Amoretti}},
  \bibinfo{author}{\bibfnamefont{R.}~\bibnamefont{Caciuffo}},
  \bibinfo{author}{\bibfnamefont{N.}~\bibnamefont{Magnani}}, \bibnamefont{and}
  \bibinfo{author}{\bibfnamefont{G.~H.} \bibnamefont{Lander}},
  \bibinfo{journal}{Rev. Mod. Phys.} \textbf{\bibinfo{volume}{81}},
  \bibinfo{pages}{807} (\bibinfo{year}{2009}).

\bibitem[{\citenamefont{Westrum~Jr. et~al.}(1953)\citenamefont{Westrum~Jr.,
  Hatcher, and Osborne}}]{westrum53}
\bibinfo{author}{\bibfnamefont{E.~F.} \bibnamefont{Westrum~Jr.}},
  \bibinfo{author}{\bibfnamefont{J.~B.} \bibnamefont{Hatcher}},
  \bibnamefont{and} \bibinfo{author}{\bibfnamefont{D.~W.}
  \bibnamefont{Osborne}}, \bibinfo{journal}{J. Chem. Phys.}
  \textbf{\bibinfo{volume}{21}}, \bibinfo{pages}{419} (\bibinfo{year}{1953}).

\bibitem[{\citenamefont{Osborne and Westrum~Jr.}(1953)}]{osborne53}
\bibinfo{author}{\bibfnamefont{D.~W.} \bibnamefont{Osborne}} \bibnamefont{and}
  \bibinfo{author}{\bibfnamefont{E.~F.} \bibnamefont{Westrum~Jr.}},
  \bibinfo{journal}{J. Chem. Phys.} \textbf{\bibinfo{volume}{21}},
  \bibinfo{pages}{1884} (\bibinfo{year}{1953}).

\bibitem[{\citenamefont{Santini et~al.}(2006)\citenamefont{Santini, Carretta,
  Magnani, Amoretti, and Caciuffo}}]{santini06}
\bibinfo{author}{\bibfnamefont{P.}~\bibnamefont{Santini}},
  \bibinfo{author}{\bibfnamefont{S.}~\bibnamefont{Carretta}},
  \bibinfo{author}{\bibfnamefont{N.}~\bibnamefont{Magnani}},
  \bibinfo{author}{\bibfnamefont{G.}~\bibnamefont{Amoretti}}, \bibnamefont{and}
  \bibinfo{author}{\bibfnamefont{R.}~\bibnamefont{Caciuffo}},
  \bibinfo{journal}{Phys. Rev. Lett.} \textbf{\bibinfo{volume}{97}},
  \bibinfo{pages}{207203} (\bibinfo{year}{2006}).

\bibitem[{\citenamefont{Magnani et~al.}(2008)\citenamefont{Magnani, Carretta,
  Caciuffo, Santini, Amoretti, Hiess, Rebizant, and Lander}}]{magnani08}
\bibinfo{author}{\bibfnamefont{N.}~\bibnamefont{Magnani}},
  \bibinfo{author}{\bibfnamefont{S.}~\bibnamefont{Carretta}},
  \bibinfo{author}{\bibfnamefont{R.}~\bibnamefont{Caciuffo}},
  \bibinfo{author}{\bibfnamefont{P.}~\bibnamefont{Santini}},
  \bibinfo{author}{\bibfnamefont{G.}~\bibnamefont{Amoretti}},
  \bibinfo{author}{\bibfnamefont{A.}~\bibnamefont{Hiess}},
  \bibinfo{author}{\bibfnamefont{J.}~\bibnamefont{Rebizant}}, \bibnamefont{and}
  \bibinfo{author}{\bibfnamefont{G.~H.} \bibnamefont{Lander}},
  \bibinfo{journal}{Phys. Rev. B} \textbf{\bibinfo{volume}{78}},
  \bibinfo{pages}{104425} (\bibinfo{year}{2008}).

\bibitem[{\citenamefont{Caciuffo et~al.}(2003)\citenamefont{Caciuffo,
  Paix{\~a}o, Detlefs, Longfield, Santini, Bernhoeft, Rebizant, and
  Lander}}]{caciuffo03}
\bibinfo{author}{\bibfnamefont{R.}~\bibnamefont{Caciuffo}},
  \bibinfo{author}{\bibfnamefont{J.~A.} \bibnamefont{Paix{\~a}o}},
  \bibinfo{author}{\bibfnamefont{C.}~\bibnamefont{Detlefs}},
  \bibinfo{author}{\bibfnamefont{M.~J.} \bibnamefont{Longfield}},
  \bibinfo{author}{\bibfnamefont{P.}~\bibnamefont{Santini}},
  \bibinfo{author}{\bibfnamefont{N.}~\bibnamefont{Bernhoeft}},
  \bibinfo{author}{\bibfnamefont{J.}~\bibnamefont{Rebizant}}, \bibnamefont{and}
  \bibinfo{author}{\bibfnamefont{G.~H.} \bibnamefont{Lander}},
  \bibinfo{journal}{J. Phys. Condens. Matter} \textbf{\bibinfo{volume}{15}},
  \bibinfo{pages}{S2287} (\bibinfo{year}{2003}).

\bibitem[{\citenamefont{Suzuki et~al.}(2010)\citenamefont{Suzuki, Magnani, and
  Oppeneer}}]{suzuki10}
\bibinfo{author}{\bibfnamefont{M.-T.} \bibnamefont{Suzuki}},
  \bibinfo{author}{\bibfnamefont{N.}~\bibnamefont{Magnani}}, \bibnamefont{and}
  \bibinfo{author}{\bibfnamefont{P.~M.} \bibnamefont{Oppeneer}},
  \bibinfo{journal}{Phys. Rev. B} \textbf{\bibinfo{volume}{82}},
  \bibinfo{pages}{241103} (\bibinfo{year}{2010}).

\bibitem[{\citenamefont{Paix\~ao et~al.}(2002)\citenamefont{Paix\~ao, Detlefs,
  Longfield, Caciuffo, Santini, Bernhoeft, Rebizant, and Lander}}]{paixao02}
\bibinfo{author}{\bibfnamefont{J.~A.} \bibnamefont{Paix\~ao}},
  \bibinfo{author}{\bibfnamefont{C.}~\bibnamefont{Detlefs}},
  \bibinfo{author}{\bibfnamefont{M.~J.} \bibnamefont{Longfield}},
  \bibinfo{author}{\bibfnamefont{R.}~\bibnamefont{Caciuffo}},
  \bibinfo{author}{\bibfnamefont{P.}~\bibnamefont{Santini}},
  \bibinfo{author}{\bibfnamefont{N.}~\bibnamefont{Bernhoeft}},
  \bibinfo{author}{\bibfnamefont{J.}~\bibnamefont{Rebizant}}, \bibnamefont{and}
  \bibinfo{author}{\bibfnamefont{G.~H.} \bibnamefont{Lander}},
  \bibinfo{journal}{Phys. Rev. Lett.} \textbf{\bibinfo{volume}{89}},
  \bibinfo{pages}{187202} (\bibinfo{year}{2002}).

\bibitem[{\citenamefont{Maehira and Hotta}(2007)}]{maehira07}
\bibinfo{author}{\bibfnamefont{T.}~\bibnamefont{Maehira}} \bibnamefont{and}
  \bibinfo{author}{\bibfnamefont{T.}~\bibnamefont{Hotta}}, \bibinfo{journal}{J.
  Magn. Magn. Mater.} \textbf{\bibinfo{volume}{310}}, \bibinfo{pages}{754}
  (\bibinfo{year}{2007}).

\bibitem[{\citenamefont{Prodan et~al.}(2007)\citenamefont{Prodan, Scuseria, and
  Martin}}]{prodan07}
\bibinfo{author}{\bibfnamefont{I.~D.} \bibnamefont{Prodan}},
  \bibinfo{author}{\bibfnamefont{G.~E.} \bibnamefont{Scuseria}},
  \bibnamefont{and} \bibinfo{author}{\bibfnamefont{R.~L.}
  \bibnamefont{Martin}}, \bibinfo{journal}{Phys. Rev. B}
  \textbf{\bibinfo{volume}{76}}, \bibinfo{pages}{033101}
  (\bibinfo{year}{2007}).

\bibitem[{\citenamefont{Wang et~al.}(2010)\citenamefont{Wang, Shi, Li, and
  Zhang}}]{wang10}
\bibinfo{author}{\bibfnamefont{B.-T.} \bibnamefont{Wang}},
  \bibinfo{author}{\bibfnamefont{H.}~\bibnamefont{Shi}},
  \bibinfo{author}{\bibfnamefont{W.}~\bibnamefont{Li}}, \bibnamefont{and}
  \bibinfo{author}{\bibfnamefont{P.}~\bibnamefont{Zhang}},
  \bibinfo{journal}{Phys. Rev. B} \textbf{\bibinfo{volume}{81}},
  \bibinfo{pages}{045119} (\bibinfo{year}{2010}).

\bibitem[{\citenamefont{Suzuki et~al.}(2013)\citenamefont{Suzuki, Magnani, and
  Oppeneer}}]{suzuki13}
\bibinfo{author}{\bibfnamefont{M.-T.} \bibnamefont{Suzuki}},
  \bibinfo{author}{\bibfnamefont{N.}~\bibnamefont{Magnani}}, \bibnamefont{and}
  \bibinfo{author}{\bibfnamefont{P.~M.} \bibnamefont{Oppeneer}},
  \bibinfo{journal}{Phys. Rev. B} \textbf{\bibinfo{volume}{88}},
  \bibinfo{pages}{195146} (\bibinfo{year}{2013}).

\bibitem[{\citenamefont{Krisch and Sette}(2007)}]{krisch07}
\bibinfo{author}{\bibfnamefont{M.}~\bibnamefont{Krisch}} \bibnamefont{and}
  \bibinfo{author}{\bibfnamefont{F.}~\bibnamefont{Sette}}, in
  \emph{\bibinfo{booktitle}{Light Scattering in Solids, Novel Materials and
  Techniques}}, edited by
  \bibinfo{editor}{\bibfnamefont{M.}~\bibnamefont{Cardona}} \bibnamefont{and}
  \bibinfo{editor}{\bibfnamefont{R.}~\bibnamefont{Merlin}}
  (\bibinfo{publisher}{Springer-Verlag, Berlin Heidelberg},
  \bibinfo{year}{2007}), vol. \bibinfo{volume}{108} of
  \emph{\bibinfo{series}{Topics in Applied Physics}}, chap.
  \bibinfo{chapter}{Inelastic X-Ray Scattering from Phonons}, pp.
  \bibinfo{pages}{317--369}.

\bibitem[{\citenamefont{Manley et~al.}(2012)\citenamefont{Manley, Jeffries,
  Said, Marianetti, Cynn, Leu, and Wall}}]{manley12}
\bibinfo{author}{\bibfnamefont{M.~E.} \bibnamefont{Manley}},
  \bibinfo{author}{\bibfnamefont{J.~R.} \bibnamefont{Jeffries}},
  \bibinfo{author}{\bibfnamefont{A.~H.} \bibnamefont{Said}},
  \bibinfo{author}{\bibfnamefont{C.~A.} \bibnamefont{Marianetti}},
  \bibinfo{author}{\bibfnamefont{H.}~\bibnamefont{Cynn}},
  \bibinfo{author}{\bibfnamefont{B.~M.} \bibnamefont{Leu}}, \bibnamefont{and}
  \bibinfo{author}{\bibfnamefont{M.~A.} \bibnamefont{Wall}},
  \bibinfo{journal}{Phys. Rev. B} \textbf{\bibinfo{volume}{85}},
  \bibinfo{pages}{132301} (\bibinfo{year}{2012}).

\bibitem[{\citenamefont{Togo et~al.}(2010)\citenamefont{Togo, Chaput, Tanaka,
  and Hug}}]{togo10}
\bibinfo{author}{\bibfnamefont{A.}~\bibnamefont{Togo}},
  \bibinfo{author}{\bibfnamefont{L.}~\bibnamefont{Chaput}},
  \bibinfo{author}{\bibfnamefont{I.}~\bibnamefont{Tanaka}}, \bibnamefont{and}
  \bibinfo{author}{\bibfnamefont{G.}~\bibnamefont{Hug}},
  \bibinfo{journal}{Phys. Rev. B} \textbf{\bibinfo{volume}{81}},
  \bibinfo{pages}{174301} (\bibinfo{year}{2010}).

\bibitem[{\citenamefont{Togo et~al.}(2015)\citenamefont{Togo, Chaput, and
  Tanaka}}]{togo15}
\bibinfo{author}{\bibfnamefont{A.}~\bibnamefont{Togo}},
  \bibinfo{author}{\bibfnamefont{L.}~\bibnamefont{Chaput}}, \bibnamefont{and}
  \bibinfo{author}{\bibfnamefont{I.}~\bibnamefont{Tanaka}},
  \bibinfo{journal}{Phys. Rev. B} \textbf{\bibinfo{volume}{91}},
  \bibinfo{pages}{094306} (\bibinfo{year}{2015}).

\bibitem[{\citenamefont{Ben-Israel and Greville}(2003)}]{adi03}
\bibinfo{author}{\bibfnamefont{A.}~\bibnamefont{Ben-Israel}} \bibnamefont{and}
  \bibinfo{author}{\bibfnamefont{T.~N.~E.} \bibnamefont{Greville}},
  \emph{\bibinfo{title}{Generalized Inverses. Theory and Applications.}}
  (\bibinfo{publisher}{Springer-Verlag, New York}, \bibinfo{year}{2003}), ISBN
  \bibinfo{isbn}{978-0-387-21634-8}.

\bibitem[{\citenamefont{Chaput}(2013)}]{chaput13}
\bibinfo{author}{\bibfnamefont{L.}~\bibnamefont{Chaput}},
  \bibinfo{journal}{Phys. Rev. Lett.} \textbf{\bibinfo{volume}{110}},
  \bibinfo{pages}{265506} (\bibinfo{year}{2013}).

\bibitem[{\citenamefont{Kresse and Furthm{\"u}ller}(1996)}]{kresse96}
\bibinfo{author}{\bibfnamefont{G.}~\bibnamefont{Kresse}} \bibnamefont{and}
  \bibinfo{author}{\bibfnamefont{J.}~\bibnamefont{Furthm{\"u}ller}},
  \bibinfo{journal}{Comput. Mater. Sci.} \textbf{\bibinfo{volume}{6}},
  \bibinfo{pages}{15} (\bibinfo{year}{1996}).

\bibitem[{\citenamefont{Perdew et~al.}(1996)\citenamefont{Perdew, Burke, and
  Ernzerhof}}]{perdew96}
\bibinfo{author}{\bibfnamefont{J.~P.} \bibnamefont{Perdew}},
  \bibinfo{author}{\bibfnamefont{K.}~\bibnamefont{Burke}}, \bibnamefont{and}
  \bibinfo{author}{\bibfnamefont{M.}~\bibnamefont{Ernzerhof}},
  \bibinfo{journal}{Phys. Rev. Lett.} \textbf{\bibinfo{volume}{77}},
  \bibinfo{pages}{3865} (\bibinfo{year}{1996}).

\bibitem[{\citenamefont{Modin et~al.}(2015)\citenamefont{Modin, Suzuki,
  Vegelius, Yun, Shuh, Werme, Nordgren, Oppeneer, and Butorin}}]{modin15}
\bibinfo{author}{\bibfnamefont{A.}~\bibnamefont{Modin}},
  \bibinfo{author}{\bibfnamefont{M.-T.} \bibnamefont{Suzuki}},
  \bibinfo{author}{\bibfnamefont{J.}~\bibnamefont{Vegelius}},
  \bibinfo{author}{\bibfnamefont{Y.}~\bibnamefont{Yun}},
  \bibinfo{author}{\bibfnamefont{D.~K.} \bibnamefont{Shuh}},
  \bibinfo{author}{\bibfnamefont{L.}~\bibnamefont{Werme}},
  \bibinfo{author}{\bibfnamefont{J.}~\bibnamefont{Nordgren}},
  \bibinfo{author}{\bibfnamefont{P.~M.} \bibnamefont{Oppeneer}},
  \bibnamefont{and} \bibinfo{author}{\bibfnamefont{S.~M.}
  \bibnamefont{Butorin}}, \bibinfo{journal}{J. Phys. Condens. Matter}
  \textbf{\bibinfo{volume}{27}}, \bibinfo{pages}{315503}
  (\bibinfo{year}{2015}).

\bibitem[{\citenamefont{Dudarev et~al.}(1997)\citenamefont{Dudarev, Manh, and
  Sutton}}]{dudarev97}
\bibinfo{author}{\bibfnamefont{S.~L.} \bibnamefont{Dudarev}},
  \bibinfo{author}{\bibfnamefont{D.~N.} \bibnamefont{Manh}}, \bibnamefont{and}
  \bibinfo{author}{\bibfnamefont{A.~P.} \bibnamefont{Sutton}},
  \bibinfo{journal}{Phil. Mag. B} \textbf{\bibinfo{volume}{75}},
  \bibinfo{pages}{613} (\bibinfo{year}{1997}).

\bibitem[{\citenamefont{Yun et~al.}(2011)\citenamefont{Yun, Rusz, Suzuki, and
  Oppeneer}}]{yun11}
\bibinfo{author}{\bibfnamefont{Y.}~\bibnamefont{Yun}},
  \bibinfo{author}{\bibfnamefont{J.}~\bibnamefont{Rusz}},
  \bibinfo{author}{\bibfnamefont{M.-T.} \bibnamefont{Suzuki}},
  \bibnamefont{and} \bibinfo{author}{\bibfnamefont{P.~M.}
  \bibnamefont{Oppeneer}}, \bibinfo{journal}{Phys. Rev. B}
  \textbf{\bibinfo{volume}{83}}, \bibinfo{pages}{075109}
  (\bibinfo{year}{2011}).

\bibitem[{\citenamefont{Togo}(2009)}]{togo09}
\bibinfo{author}{\bibfnamefont{A.}~\bibnamefont{Togo}},
  \emph{\bibinfo{title}{Phonopy}} (\bibinfo{year}{2009}),
  \urlprefix\url{http://phonopy.sourceforge.net/index.html}.

\bibitem[{\citenamefont{Pick et~al.}(1970)\citenamefont{Pick, Cohen, and
  Martin}}]{pick70}
\bibinfo{author}{\bibfnamefont{R.~M.} \bibnamefont{Pick}},
  \bibinfo{author}{\bibfnamefont{M.~H.} \bibnamefont{Cohen}}, \bibnamefont{and}
  \bibinfo{author}{\bibfnamefont{R.~M.} \bibnamefont{Martin}},
  \bibinfo{journal}{Phys. Rev. B} \textbf{\bibinfo{volume}{1}},
  \bibinfo{pages}{910} (\bibinfo{year}{1970}).

\bibitem[{\citenamefont{Sanati et~al.}(2011)\citenamefont{Sanati, Albers,
  Lookman, and Saxena}}]{sanati11}
\bibinfo{author}{\bibfnamefont{M.}~\bibnamefont{Sanati}},
  \bibinfo{author}{\bibfnamefont{R.~C.} \bibnamefont{Albers}},
  \bibinfo{author}{\bibfnamefont{T.}~\bibnamefont{Lookman}}, \bibnamefont{and}
  \bibinfo{author}{\bibfnamefont{A.}~\bibnamefont{Saxena}},
  \bibinfo{journal}{Phys. Rev. B} \textbf{\bibinfo{volume}{84}},
  \bibinfo{pages}{014116} (\bibinfo{year}{2011}).

\bibitem[{\citenamefont{Dorado et~al.}(2009)\citenamefont{Dorado, Amadon,
  Freyss, and Bertolus}}]{dorado09}
\bibinfo{author}{\bibfnamefont{B.}~\bibnamefont{Dorado}},
  \bibinfo{author}{\bibfnamefont{B.}~\bibnamefont{Amadon}},
  \bibinfo{author}{\bibfnamefont{M.}~\bibnamefont{Freyss}}, \bibnamefont{and}
  \bibinfo{author}{\bibfnamefont{M.}~\bibnamefont{Bertolus}},
  \bibinfo{journal}{Phys. Rev. B} \textbf{\bibinfo{volume}{79}},
  \bibinfo{pages}{235125} (\bibinfo{year}{2009}).

\bibitem[{\citenamefont{Kune\ifmmode~\check{s}\else \v{s}\fi{}
  et~al.}(2001)\citenamefont{Kune\ifmmode~\check{s}\else \v{s}\fi{}, Nov\'ak,
  Divi\ifmmode~\check{s}\else \v{s}\fi{}, and Oppeneer}}]{kunes01}
\bibinfo{author}{\bibfnamefont{J.}~\bibnamefont{Kune\ifmmode~\check{s}\else
  \v{s}\fi{}}}, \bibinfo{author}{\bibfnamefont{P.}~\bibnamefont{Nov\'ak}},
  \bibinfo{author}{\bibfnamefont{M.}~\bibnamefont{Divi\ifmmode~\check{s}\else
  \v{s}\fi{}}}, \bibnamefont{and} \bibinfo{author}{\bibfnamefont{P.~M.}
  \bibnamefont{Oppeneer}}, \bibinfo{journal}{Phys. Rev. B}
  \textbf{\bibinfo{volume}{63}}, \bibinfo{pages}{205111}
  (\bibinfo{year}{2001}).

\bibitem[{\citenamefont{Boettger and Ray}(2000)}]{boettger00}
\bibinfo{author}{\bibfnamefont{J.~C.} \bibnamefont{Boettger}} \bibnamefont{and}
  \bibinfo{author}{\bibfnamefont{A.~K.} \bibnamefont{Ray}},
  \bibinfo{journal}{Int. J. Quantum Chem.} \textbf{\bibinfo{volume}{80}},
  \bibinfo{pages}{824} (\bibinfo{year}{2000}).

\bibitem[{\citenamefont{Prodan et~al.}(2006)\citenamefont{Prodan, Scuseria, and
  Martin}}]{prodan06}
\bibinfo{author}{\bibfnamefont{I.~D.} \bibnamefont{Prodan}},
  \bibinfo{author}{\bibfnamefont{G.~E.} \bibnamefont{Scuseria}},
  \bibnamefont{and} \bibinfo{author}{\bibfnamefont{R.~L.}
  \bibnamefont{Martin}}, \bibinfo{journal}{Phys. Rev. B}
  \textbf{\bibinfo{volume}{73}}, \bibinfo{pages}{045104}
  (\bibinfo{year}{2006}).

\bibitem[{\citenamefont{Petit et~al.}(2010)\citenamefont{Petit, Svane, Szotek,
  Temmerman, and Stocks}}]{petit10}
\bibinfo{author}{\bibfnamefont{L.}~\bibnamefont{Petit}},
  \bibinfo{author}{\bibfnamefont{A.}~\bibnamefont{Svane}},
  \bibinfo{author}{\bibfnamefont{Z.}~\bibnamefont{Szotek}},
  \bibinfo{author}{\bibfnamefont{W.~M.} \bibnamefont{Temmerman}},
  \bibnamefont{and} \bibinfo{author}{\bibfnamefont{G.~M.}
  \bibnamefont{Stocks}}, \bibinfo{journal}{Phys. Rev. B}
  \textbf{\bibinfo{volume}{81}}, \bibinfo{pages}{045108}
  (\bibinfo{year}{2010}).

\bibitem[{\citenamefont{Elliott and Thorpe}(1967)}]{elliott67}
\bibinfo{author}{\bibfnamefont{R.~J.} \bibnamefont{Elliott}} \bibnamefont{and}
  \bibinfo{author}{\bibfnamefont{M.~F.} \bibnamefont{Thorpe}},
  \bibinfo{journal}{Proc. Phys. Soc.} \textbf{\bibinfo{volume}{91}},
  \bibinfo{pages}{903} (\bibinfo{year}{1967}).

\bibitem[{\citenamefont{Serizawa et~al.}(2001)\citenamefont{Serizawa, Arai, and
  Nakajima}}]{serizawa01}
\bibinfo{author}{\bibfnamefont{H.}~\bibnamefont{Serizawa}},
  \bibinfo{author}{\bibfnamefont{Y.}~\bibnamefont{Arai}}, \bibnamefont{and}
  \bibinfo{author}{\bibfnamefont{K.}~\bibnamefont{Nakajima}},
  \bibinfo{journal}{J. Chem. Thermodyn.} \textbf{\bibinfo{volume}{33}},
  \bibinfo{pages}{615} (\bibinfo{year}{2001}).

\bibitem[{\citenamefont{Fournier et~al.}(1991)\citenamefont{Fournier, Blaise,
  Amoretti, Caciuffo, Larroque, Hutchings, Osborn, and Taylor}}]{fournier91}
\bibinfo{author}{\bibfnamefont{J.~M.} \bibnamefont{Fournier}},
  \bibinfo{author}{\bibfnamefont{A.}~\bibnamefont{Blaise}},
  \bibinfo{author}{\bibfnamefont{G.}~\bibnamefont{Amoretti}},
  \bibinfo{author}{\bibfnamefont{R.}~\bibnamefont{Caciuffo}},
  \bibinfo{author}{\bibfnamefont{J.}~\bibnamefont{Larroque}},
  \bibinfo{author}{\bibfnamefont{M.~T.} \bibnamefont{Hutchings}},
  \bibinfo{author}{\bibfnamefont{R.}~\bibnamefont{Osborn}}, \bibnamefont{and}
  \bibinfo{author}{\bibfnamefont{A.~D.} \bibnamefont{Taylor}},
  \bibinfo{journal}{Phys. Rev. B} \textbf{\bibinfo{volume}{43}},
  \bibinfo{pages}{1142} (\bibinfo{year}{1991}).

\bibitem[{\citenamefont{Caciuffo et~al.}(1992)\citenamefont{Caciuffo, Amoretti,
  Blaise, Fournier, Hutchings, Lander, Osborn, Severing, and
  Taylor}}]{caciuffo92}
\bibinfo{author}{\bibfnamefont{R.}~\bibnamefont{Caciuffo}},
  \bibinfo{author}{\bibfnamefont{G.}~\bibnamefont{Amoretti}},
  \bibinfo{author}{\bibfnamefont{A.}~\bibnamefont{Blaise}},
  \bibinfo{author}{\bibfnamefont{J.~M.} \bibnamefont{Fournier}},
  \bibinfo{author}{\bibfnamefont{M.~T.} \bibnamefont{Hutchings}},
  \bibinfo{author}{\bibfnamefont{G.~H.} \bibnamefont{Lander}},
  \bibinfo{author}{\bibfnamefont{R.}~\bibnamefont{Osborn}},
  \bibinfo{author}{\bibfnamefont{A.}~\bibnamefont{Severing}}, \bibnamefont{and}
  \bibinfo{author}{\bibfnamefont{A.~D.} \bibnamefont{Taylor}},
  \bibinfo{journal}{Physica B: Condens. Matter} \textbf{\bibinfo{volume}{180}},
  \bibinfo{pages}{149} (\bibinfo{year}{1992}).

\bibitem[{\citenamefont{Magnani
  et~al.}(2005{\natexlab{b}})\citenamefont{Magnani, Santini, Amoretti,
  Caciuffo, Javorsk\'y, Wastin, Rebizant, and Lander}}]{magnani05a}
\bibinfo{author}{\bibfnamefont{N.}~\bibnamefont{Magnani}},
  \bibinfo{author}{\bibfnamefont{P.}~\bibnamefont{Santini}},
  \bibinfo{author}{\bibfnamefont{G.}~\bibnamefont{Amoretti}},
  \bibinfo{author}{\bibfnamefont{R.}~\bibnamefont{Caciuffo}},
  \bibinfo{author}{\bibfnamefont{P.}~\bibnamefont{Javorsk\'y}},
  \bibinfo{author}{\bibfnamefont{F.}~\bibnamefont{Wastin}},
  \bibinfo{author}{\bibfnamefont{J.}~\bibnamefont{Rebizant}}, \bibnamefont{and}
  \bibinfo{author}{\bibfnamefont{G.~H.} \bibnamefont{Lander}},
  \bibinfo{journal}{Physica B} \textbf{\bibinfo{volume}{359-361}},
  \bibinfo{pages}{1087} (\bibinfo{year}{2005}{\natexlab{b}}).

\bibitem[{\citenamefont{Bene$\check{s}$
  et~al.}(2011)\citenamefont{Bene$\check{s}$, Gotcu-Freis, Schw{\"o}rer,
  Konings, and Fangh\"{a}nel}}]{benes11}
\bibinfo{author}{\bibfnamefont{O.}~\bibnamefont{Bene$\check{s}$}},
  \bibinfo{author}{\bibfnamefont{P.}~\bibnamefont{Gotcu-Freis}},
  \bibinfo{author}{\bibfnamefont{F.}~\bibnamefont{Schw{\"o}rer}},
  \bibinfo{author}{\bibfnamefont{R.~J.~M.} \bibnamefont{Konings}},
  \bibnamefont{and}
  \bibinfo{author}{\bibfnamefont{T.}~\bibnamefont{Fangh\"{a}nel}},
  \bibinfo{journal}{J. Chem. Thermodyn.} \textbf{\bibinfo{volume}{43}},
  \bibinfo{pages}{651 } (\bibinfo{year}{2011}).

\bibitem[{\citenamefont{Yun et~al.}(2012)\citenamefont{Yun, Legut, and
  Oppeneer}}]{yun12}
\bibinfo{author}{\bibfnamefont{Y.}~\bibnamefont{Yun}},
  \bibinfo{author}{\bibfnamefont{D.}~\bibnamefont{Legut}}, \bibnamefont{and}
  \bibinfo{author}{\bibfnamefont{P.~M.} \bibnamefont{Oppeneer}},
  \bibinfo{journal}{J. Nucl. Mater.} \textbf{\bibinfo{volume}{426}},
  \bibinfo{pages}{109} (\bibinfo{year}{2012}).

\bibitem[{\citenamefont{Benedict et~al.}(1986)\citenamefont{Benedict, Dabos,
  Dufour, Spirlet, and Pag\`{e}s}}]{benedict86}
\bibinfo{author}{\bibfnamefont{U.}~\bibnamefont{Benedict}},
  \bibinfo{author}{\bibfnamefont{S.}~\bibnamefont{Dabos}},
  \bibinfo{author}{\bibfnamefont{C.}~\bibnamefont{Dufour}},
  \bibinfo{author}{\bibfnamefont{J.~C.} \bibnamefont{Spirlet}},
  \bibnamefont{and}
  \bibinfo{author}{\bibfnamefont{M.}~\bibnamefont{Pag\`{e}s}},
  \bibinfo{journal}{J. Less-Comm. Metals} \textbf{\bibinfo{volume}{121}},
  \bibinfo{pages}{461 } (\bibinfo{year}{1986}).

\bibitem[{\citenamefont{Sobolev}(2009)}]{sobolev09}
\bibinfo{author}{\bibfnamefont{V.}~\bibnamefont{Sobolev}}, \bibinfo{journal}{J.
  Nucl. Mater.} \textbf{\bibinfo{volume}{389}}, \bibinfo{pages}{45 }
  (\bibinfo{year}{2009}).

\bibitem[{\citenamefont{Gofryk et~al.}(2014)\citenamefont{Gofryk, Du, Stanek,
  Lashley, Liu, Schulze, Smith, Safarik, Byler, McClellan et~al.}}]{gofryk14}
\bibinfo{author}{\bibfnamefont{K.}~\bibnamefont{Gofryk}},
  \bibinfo{author}{\bibfnamefont{S.}~\bibnamefont{Du}},
  \bibinfo{author}{\bibfnamefont{C.~R.} \bibnamefont{Stanek}},
  \bibinfo{author}{\bibfnamefont{J.~C.} \bibnamefont{Lashley}},
  \bibinfo{author}{\bibfnamefont{X.~Y.} \bibnamefont{Liu}},
  \bibinfo{author}{\bibfnamefont{R.~K.} \bibnamefont{Schulze}},
  \bibinfo{author}{\bibfnamefont{J.~L.} \bibnamefont{Smith}},
  \bibinfo{author}{\bibfnamefont{D.~J.} \bibnamefont{Safarik}},
  \bibinfo{author}{\bibfnamefont{D.~D.} \bibnamefont{Byler}},
  \bibinfo{author}{\bibfnamefont{K.~J.} \bibnamefont{McClellan}},
 \bibinfo{author}{\bibfnamefont{B.~P.} \bibnamefont{Uberuaga}},
 \bibinfo{author}{\bibfnamefont{B.~L.} \bibnamefont{Scott}},
  \bibinfo{author}{\bibfnamefont{D.~A.} \bibnamefont{Andersson}}, \bibinfo{journal}{Nature Commun.}
  \textbf{\bibinfo{volume}{5}}, \bibinfo{pages}{4551} (\bibinfo{year}{2014}).

\bibitem[{\citenamefont{Nishi et~al.}(2008)\citenamefont{Nishi, Itoh, Takano,
  Numata, Akabori, Arai, and Minato}}]{nishi08}
\bibinfo{author}{\bibfnamefont{T.}~\bibnamefont{Nishi}},
  \bibinfo{author}{\bibfnamefont{A.}~\bibnamefont{Itoh}},
  \bibinfo{author}{\bibfnamefont{M.}~\bibnamefont{Takano}},
  \bibinfo{author}{\bibfnamefont{M.}~\bibnamefont{Numata}},
  \bibinfo{author}{\bibfnamefont{M.}~\bibnamefont{Akabori}},
  \bibinfo{author}{\bibfnamefont{Y.}~\bibnamefont{Arai}}, \bibnamefont{and}
  \bibinfo{author}{\bibfnamefont{K.}~\bibnamefont{Minato}},
  \bibinfo{journal}{J. Nucl. Mater.} \textbf{\bibinfo{volume}{376}},
  \bibinfo{pages}{78 } (\bibinfo{year}{2008}).

\bibitem[{\citenamefont{Minato et~al.}(2009)\citenamefont{Minato, Takano,
  Otobe, Nishi, Akabori, and Arai}}]{minato09}
\bibinfo{author}{\bibfnamefont{K.}~\bibnamefont{Minato}},
  \bibinfo{author}{\bibfnamefont{M.}~\bibnamefont{Takano}},
  \bibinfo{author}{\bibfnamefont{H.}~\bibnamefont{Otobe}},
  \bibinfo{author}{\bibfnamefont{T.}~\bibnamefont{Nishi}},
  \bibinfo{author}{\bibfnamefont{M.}~\bibnamefont{Akabori}}, \bibnamefont{and}
  \bibinfo{author}{\bibfnamefont{Y.}~\bibnamefont{Arai}}, \bibinfo{journal}{J.
  Nucl. Mater.} \textbf{\bibinfo{volume}{389}}, \bibinfo{pages}{23 }
  (\bibinfo{year}{2009}).

\bibitem[{\citenamefont{Yin and Savrasov}(2008)}]{yin08}
\bibinfo{author}{\bibfnamefont{Q.}~\bibnamefont{Yin}} \bibnamefont{and}
  \bibinfo{author}{\bibfnamefont{S.~Y.} \bibnamefont{Savrasov}},
  \bibinfo{journal}{Phys. Rev. Lett.} \textbf{\bibinfo{volume}{100}},
  \bibinfo{pages}{225504} (\bibinfo{year}{2008}).

\bibitem[{\citenamefont{Hyland}(1983)}]{hyland83}
\bibinfo{author}{\bibfnamefont{G.~J.} \bibnamefont{Hyland}},
  \bibinfo{journal}{J. Nucl. Mater.} \textbf{\bibinfo{volume}{113}},
  \bibinfo{pages}{125} (\bibinfo{year}{1983}).

\bibitem[{\citenamefont{Fink}(2000)}]{fink00}
\bibinfo{author}{\bibfnamefont{J.~K.} \bibnamefont{Fink}}, \bibinfo{journal}{J.
  Nucl. Mater.} \textbf{\bibinfo{volume}{279}}, \bibinfo{pages}{1}
  (\bibinfo{year}{2000}).

\bibitem[{\citenamefont{Idiri et~al.}(2004)\citenamefont{Idiri, Le~Bihan,
  Heathman, and Rebizant}}]{idiri04}
\bibinfo{author}{\bibfnamefont{M.}~\bibnamefont{Idiri}},
  \bibinfo{author}{\bibfnamefont{T.}~\bibnamefont{Le~Bihan}},
  \bibinfo{author}{\bibfnamefont{S.}~\bibnamefont{Heathman}}, \bibnamefont{and}
  \bibinfo{author}{\bibfnamefont{J.}~\bibnamefont{Rebizant}},
  \bibinfo{journal}{Phys. Rev. B} \textbf{\bibinfo{volume}{70}},
  \bibinfo{pages}{014113} (\bibinfo{year}{2004}).

\end{thebibliography}

\end{document}